%% file: icml2024/main.tex
\documentclass{article}

\usepackage{microtype}
\usepackage{graphicx}
\usepackage{subcaption}
\usepackage{booktabs} % for professional tables
\usepackage{multirow}
\usepackage{makecell}
\usepackage{tablefootnote}
\usepackage{pifont}
\usepackage[dvipsnames]{xcolor}

\usepackage{hyperref}

\usepackage[accepted]{icml2024}

\usepackage{amsmath}
\usepackage{amssymb}
\usepackage{mathtools}
\usepackage{amsthm}

\usepackage[capitalize,noabbrev]{cleveref}

\theoremstyle{plain}

\theoremstyle{definition}

\theoremstyle{remark}

\usepackage[textsize=tiny]{todonotes}

\icmltitlerunning{Generative Audio Language Modeling with Continuous-valued Tokens and Masked Next-Token Prediction}

\begin{document}

\twocolumn[
\icmltitle{Generative Audio Language Modeling with Continuous-valued Tokens and Masked Next-Token Prediction}



\icmlsetsymbol{equal}{*}

\begin{icmlauthorlist}
\icmlauthor{Shu-wen Yang}{ntu,amazon}
\icmlauthor{Byeonggeun Kim}{equal,amazon}
\icmlauthor{Kuan-Po Huang}{equal,ntu,amazon}
\icmlauthor{Qingming Tang}{amazon}
\icmlauthor{Huy Phan}{amazon}
\icmlauthor{Bo-Ru Lu}{amazon}
\icmlauthor{Harsha Sundar}{amazon}
\icmlauthor{Shalini Ghosh}{amazon}
\icmlauthor{Hung-yi Lee}{ntu}
\icmlauthor{Chieh-Chi Kao}{amazon}
\icmlauthor{Chao Wang}{amazon}
\end{icmlauthorlist}

\icmlaffiliation{ntu}{Graduate Institute of Communication Engineering, National Taiwan University, Taipei, Taiwan}
\icmlaffiliation{amazon}{Amazon AGI, Bellevue, United States}

\icmlcorrespondingauthor{Shu-wen Yang}{leo19941227@gmail.com}
\icmlcorrespondingauthor{Chieh-Chi Kao}{chiehchi@amazon.com}

\icmlkeywords{Machine Learning, ICML}

\vskip 0.3in
]



\printAffiliationsAndNotice{\icmlEqualContribution} % otherwise use the standard text.

\input{icml2024/sections/abstract}
\input{icml2024/sections/introduction}
\input{icml2024/sections/related_work}
\input{icml2024/sections/method}
\input{icml2024/sections/implementation}
\input{icml2024/sections/experiments}
\input{icml2024/sections/conclusion}

% Authors are kindly asked to make their submissions as accessible as possible for everyone including people with disabilities and sensory or neurological differences.
% Tips of how to achieve this and what to pay attention to will be provided on the conference website \url{http://icml.cc/}.

% \section*{Software and Data}

% If a paper is accepted, we strongly encourage the publication of software and data with the
% camera-ready version of the paper whenever appropriate. This can be
% done by including a URL in the camera-ready copy. However, \textbf{do not}
% include URLs that reveal your institution or identity in your
% submission for review. Instead, provide an anonymous URL or upload
% the material as ``Supplementary Material'' into the OpenReview reviewing
% system. Note that reviewers are not required to look at this material
% when writing their review.

% \section*{Acknowledgements}

% \textbf{Do not} include acknowledgements in the initial version of
% the paper submitted for blind review.

% If a paper is accepted, the final camera-ready version can (and
% usually should) include acknowledgements.  Such acknowledgements
% should be placed at the end of the section, in an unnumbered section
% that does not count towards the paper page limit. Typically, this will 
% include thanks to reviewers who gave useful comments, to colleagues 
% who contributed to the ideas, and to funding agencies and corporate 
% sponsors that provided financial support.

\section*{Impact Statement}

\paragraph{Scaling with Large Language Models (LLMs).}

Our work demonstrates that causal language modeling can produce high-fidelity audio even with small models and limited data. A natural next step is to scale up to billions of parameters like LLMs and incorporate larger datasets, such as AudioSet\footnote{We currently rely only on AudioCaps and WavCaps, totaling about 1,000 hours of audio.}. With this scaling, our approach has the potential to set a new SOTA in streamable sound generation.

\paragraph{Merge with other Modalities in LLMs.}

Since our method only requires changing the prediction head of the Transformer decoder, it can be easily integrated into existing multi-modality LLM training to jointly learn with other modalities such as text and image~\cite{sun2024multimodal}.
Compared to post-synthesizing with the external vocoders~\cite{huang2024audiogpt,liu2023wavjourney,wang2024audio}, the end-to-end approach avoids error propagation and allows audio data to interact directly and losslessly with other modalities in a single model.
Audio generation also benefits from the LLM scaling with other modalities.

\paragraph{Efficient Inference.}

Since we primarily adhere to the standard Transformer decoder, our model naturally benefits from resources developed for this standardized architecture, including the KV-cache~\cite{pope2023efficiently} and PagedAttention~\cite{kwon2023efficient}.
Furthermore, our model learns to \textit{predict any future timestamp given any subset of past information}.
Theoretically, one can devise a parallel decoding algorithm by manipulating the target positional embedding.
With these techniques, combined with our inherently streaming capability, our method can potentially establish a new SOTA in efficient, high-fidelity sound generation.


\bibliography{ref}
\bibliographystyle{icml2024}

%%%%%%%%%%%%%%%%%%%%%%%%%%%%%%%%%%%%%%%%%%%%%%%%%%%%%%%%%%%%%%%%%%%%%%%%%%%%%%%
%%%%%%%%%%%%%%%%%%%%%%%%%%%%%%%%%%%%%%%%%%%%%%%%%%%%%%%%%%%%%%%%%%%%%%%%%%%%%%%
% APPENDIX
%%%%%%%%%%%%%%%%%%%%%%%%%%%%%%%%%%%%%%%%%%%%%%%%%%%%%%%%%%%%%%%%%%%%%%%%%%%%%%%
%%%%%%%%%%%%%%%%%%%%%%%%%%%%%%%%%%%%%%%%%%%%%%%%%%%%%%%%%%%%%%%%%%%%%%%%%%%%%%%
\newpage
\appendix
\onecolumn

\input{icml2024/sections/appendix}

%%%%%%%%%%%%%%%%%%%%%%%%%%%%%%%%%%%%%%%%%%%%%%%%%%%%%%%%%%%%%%%%%%%%%%%%%%%%%%%
%%%%%%%%%%%%%%%%%%%%%%%%%%%%%%%%%%%%%%%%%%%%%%%%%%%%%%%%%%%%%%%%%%%%%%%%%%%%%%%

\end{document}

%% file: icml2024/sections/abstract.tex
\begin{abstract}

Autoregressive next-token prediction with the Transformer decoder has become a de facto standard in large language models (LLMs), achieving remarkable success in Natural Language Processing (NLP) at scale.
Extending this paradigm to audio poses unique challenges due to its inherently continuous nature.
% unlike discrete text.
We research audio generation with a causal language model (LM) without discrete tokens.
We leverage token-wise diffusion to model the continuous distribution of the next continuous-valued token. Our approach delivers significant improvements over previous discrete solution, AudioGen, achieving 20\% and 40\% relative gains on AudioCaps in Frechet Audio Distance (FAD) and Kullback-Leibler (KL) divergence, respectively.
Additionally, we propose a novel masked next-token prediction task that incorporates masked prediction into the causal LM framework.
On AudioCaps, the innovation yields 41\% and 33\% relative FAD improvements over AudioGen Base (285M) and AudioGen Large (1B) models, respectively, and is on par with the state-of-the-art (SOTA) diffusion models.
Furthermore, we achieve these results with significantly fewer parameters—193M for our Base and 462M for our Large models.
% Our results highlight the promising potential of directly generating high-fidelity audio with LLMs.

\end{abstract}

%% file: icml2024/sections/introduction.tex
\section{Introduction}
\label{sec:introduction}

\begin{figure}[th]
% \vskip 0.1in
\begin{center}
\centerline{\includegraphics[width=1\columnwidth]{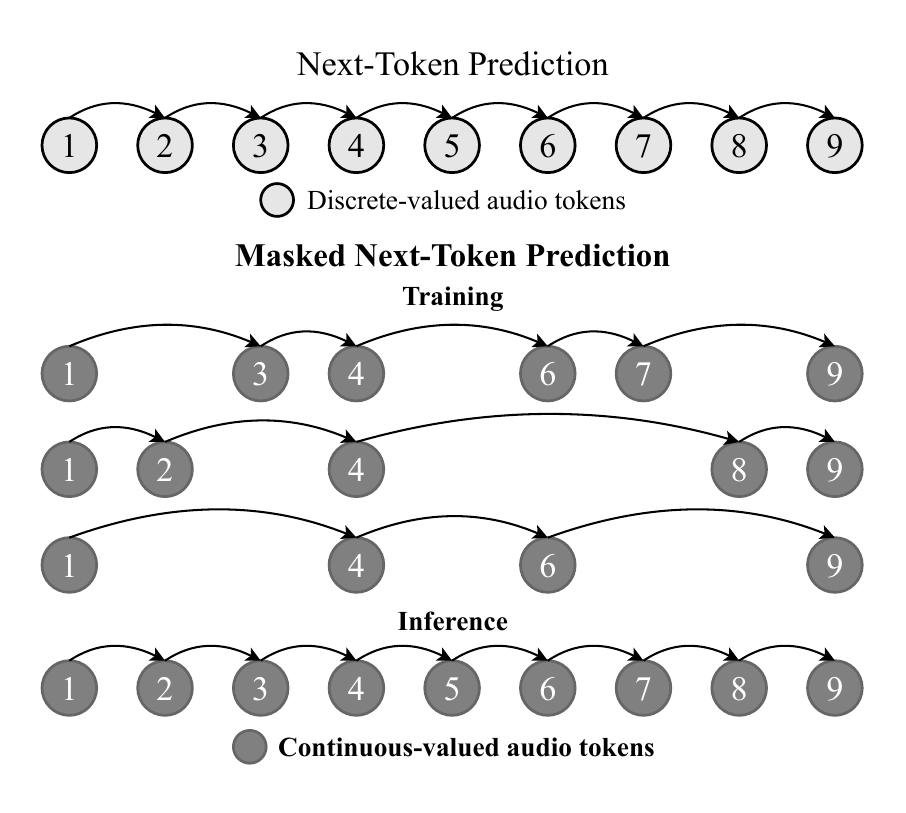}}
\caption{\textbf{Causal Language Modeling on Continuous-valued Audio Tokens with Masked Next-Token Prediction.} The audio tokens are low-dimensional continuous latent.
We use a standard Transformer decoder for the audio language modeling.
Our masked next-token prediction learns to predict any future token given any subset of the past tokens, which turns out benefit the standard next-token prediction and rival the bidirectional diffusion models and masked generative models (i.e. MAR~\cite{li2024autoregressive}).
The figure illustrates the prediction order in which all tokens are conditioned on all previously predicted tokens.
}
\label{image:front}
\end{center}
\vskip -0.3in
\end{figure}

% Why decoder-only model is so important
Large Language Models (LLMs) has revolutionized artificial intelligence~\cite{glue,super-glue,mmlu}.
They show remarkable emergent capabilities and human-level understanding and reasoning abilities~\cite{gpt4,team2023gemini} after scaling up to thousands of billions of parameters, such as the GPT series~\cite{brown2020language,gpt4}.
The essence behind LLMs is simple, \textit{scaling up the model by more parameters and data with a Transformer decoder~\cite{vaswani2017attention} and next-token prediction}.
% By learning to predict future in more diverse conditions, the model achieves higher level of intelligence.
Given the success of this paradigm, plenty of efforts have been put into developing better theories and tools for this standardized model, including the architecture improvement~\cite{chowdhery2023palm,du2022glam} scaling laws~\cite{kaplan2020scaling,hoffmann2022training}, and efficient inference and streaming infrastructure~\cite{kwon2023efficient,pope2023efficiently}.

% Why audio generation and why using decoder-only model in audio generation
We study text-to-audio (TTA) with the LLM framework. TTA is critical due to its various applications in virtual (augmented) reality, media creation, video editing, and game development~\cite{liu2023audioldm}. Specifically, given a natural language prompt, the model generates a relevant sound.
The causal language modeling approach is important to audio for its fundamental role in multi-modal LLMs~\cite{wunext,yin2023survey,liu2024visual,zhou2024transfusion} and the streaming capability for real-time user interaction~\cite{wang2024freeze,zhang2024intrinsicvoice}. Hence, we ask \textit{whether a causal language model can achieve high-fidelity audio generation.}
Existing works applied a Transformer decoder with next-token prediction on discrete audio tokens for TTA~\cite{AudioGen,uniaudio}.
Despite scaling to billions of parameters, these models lag behind the SOTA systems, which typically rely on diffusion techniques~\cite{tango2,AudioLDM2}.
As a result, to build an LLM that can speak or produce high-fidelity sound, the prevailing approach relies on external vocoders~\cite{huang2024audiogpt,liu2023wavjourney,wang2024audio}. However, the pipeline approach overlooks the possibility of directly generating audio with a scalable and streamable LLM, allowing us to enjoy the aforementioned resources like scaling, efficient inference and streaming infrastructure.

% continuous-valued tokens
In this work, we enable audio generation for causal language models.
We address the bottlenecks that block the decoder-only solution by integrating strengths from SOTA diffusion models~\cite{tango2,AudioLDM2}, including token types, loss functions, and learning tasks.
We reinterpret the variational auto-encoder (VAE) latents in latent diffusion models (LDMs)~\cite{rombach2022high} as \textit{continuous-valued tokens}\footnote{
Specifically, we use a VAE to encode the Mel-spectrogram into a 2-D feature map, and serialize it into a 1-D sequence of low-dimensional latents.
}, replacing discrete acoustic tokens used in audio language modeling. 
To model the continuous distribution of the next token, we replace the cross-entropy loss by the token-wise diffusion loss~\cite{li2024autoregressive}, leaving the backbone Transformer decoder intact,  allowing it to fully leverage the benefits and resources of LLMs.
Using continuous-valued \textbf{audio} tokens and \textbf{next-token prediction (NTP)}, our 193M \textbf{AudioNTP Base} achieves 20\% and 40\% relative improvements over 285M AudioGen Base~\cite{AudioGen} on AudioCaps~\cite{kim2019audiocaps} in Frechet Audio Distance (FAD) and Kullback-Leibler (KL) divergence, respectively.

% MNTP
To further match the performance of SOTA diffusion models, we incorporate masked language modeling (MLM) and propose a novel learning task.
MLM learns contextualized dependencies~\cite{kenton2019bert,vyas2023audiobox,liugenerative,li2024autoregressive,chang2022maskgit} and develops understanding capabilities~\cite{li2024return,li2023mage,wei2023diffusion}, both of which have been shown to benefit generation across various scenarios.
Concretely, we randomly drop tokens as the masking, forming a shorter sequence.
Next-token prediction is then performed on this shorter sequence as the masked prediction, a task we term \textbf{masked next-token prediction (MNTP)}.
Compared to next-token prediction, MNTP predicts a random future token conditioned on a random subset of past tokens (Figure~\ref{image:front}), a task that ultimately benefits next-token prediction.
On AudioCaps, our \textbf{AudioMNTP Base} significantly outperforms AudioNTP Base, 
achieving a 41\% relative FAD improvement over AudioGen Base.
Scaling up to 462M \textbf{AudioMNTP Large} yields a 33\% relative FAD improvement 
over 1B AudioGen Large, matching SOTA diffusion models~\cite{AudioLDM2,tango2} 
while remaining streamable and compatible with LLMs.
Our contributions are:

\begin{itemize}
    \item We propose the use of continuous-valued tokens in generative audio language modeling, demonstrating their superiority over discrete tokens.
    \item We introduce a novel learning task, masked next-token prediction (MNTP). Training with MNTP enhances the model's decoding performance on the regular next-token prediction.
    \item By integrating continuous-valued tokens and MNTP, we achieve SOTA-level audio generation within the causal LM framework, establishing a pathway for directly generating high-fidelity audio with LLMs.
\end{itemize}

%% file: icml2024/sections/related_work.tex
\section{Related Work}

\paragraph{Text-guided Audio Generation.}

The most relevant works include AudioGen~\cite{AudioGen} and UniAudio~\cite{uniaudio}. AudioGen learns a causal LM on discrete tokens~\cite{zeghidour2021soundstream,defossezhigh} with low downsampling rate to preserve audio fidelity, whereas UniAudio employs a multi-scale Transformer~\cite{yu2023megabyte} to reduce RVQ token sequence length, thereby saving computation. On the other hand, LDMs~\cite{rombach2022high} have outperformed language modeling approaches in audio generation~\cite{yang2023diffsound,AudioLDM2,tango2} by modeling the joint probability distribution of continuous-valued tokens through diffusion processes.
Among them, we adopt a causal LM similar to AudioGen but apply it to continuous-valued tokens, as in LDMs.

\paragraph{Language Modeling on Continuous-valued Tokens.}

Language modeling on continuous data has traditionally required quantizing data into discrete tokens~\cite{chen2020generative,esser2021taming,borsos2023audiolm,AudioGen}. Recent works in image and speech generation~\cite{li2024autoregressive,tschannen2025givt,meng2024autoregressive} demonstrates that continuous-valued tokens prevent information loss and achieve higher generation quality.
These approaches model continuous next-token distributions using diffusion loss~\cite{li2024autoregressive} or Gaussian mixture models~\cite{tschannen2025givt,meng2024autoregressive}.
In our work, we employ diffusion loss for sound.

\paragraph{Masked Prediction in Generative Models.}

Masked prediction has been shown effective in various generative models. In text generation, masked infilling serves as a pretext task for bidirectional models~\cite{raffel2020exploring,lewis-etal-2020-bart}, and iterative MLM has been applied to image generation~\cite{li2023mage,chang2022maskgit}. Similarly, masked infilling benefits audio~\cite{vyas2023audiobox} and speech~\cite{liugenerative} generation in the flow-matching~\cite{lipmanflow} framework.
Recently, several methods~\cite{friedincoder,Aghajanyan2022CM3AC,peng-etal-2024-voicecraft,bavarian2022efficient} reorder masked tokens to the sequence end, enabling decoder-only models to leverage both past and future context for the editing tasks. In contrast, we integrate masked prediction without future context, strictly preserving causality and improving unidirectional decoding.

% emphasize our difference and contribution upon them
\paragraph{Relation to MAR.}
\label{section:relation_to_mar}

Our work is based on MAR~\cite{li2024autoregressive}, which introduces continuous-valued tokens in the context of masked generative modeling (MGM) for image generation. In MGM, a bidirectional model iteratively performs MLM from an all-zero input, predicting all masked positions but retaining only a subset, until all positions are retained.
Our approach differs from MAR by using a standard Transformer decoder with the classic next-token prediction, making it compatible with LLMs. We show that, in the audio domain using continuous-valued tokens, the unidirectional MNTP outperforms naive next-token prediction and is comparable to the bidirectional MAR.
Furthermore, in the left-to-right causal inference scenario, MNTP outperforms MAR.

%% file: icml2024/sections/method.tex
\section{Method}

\newcommand{\e}{\varepsilon}
\newcommand{\abar}{\bar{\alpha}}

Our approach includes two main proposals: continuous-valued tokens in Section~\ref{section:AudioNTP} and masked next-token prediction in Section~\ref{section:AudioMNTP}.

\subsection{AudioNTP: Continuous-valued Tokens}
\label{section:AudioNTP}

\begin{figure}[t]
% \vskip 0.2in
\begin{center}
\centerline{\includegraphics[width=1.1\columnwidth]{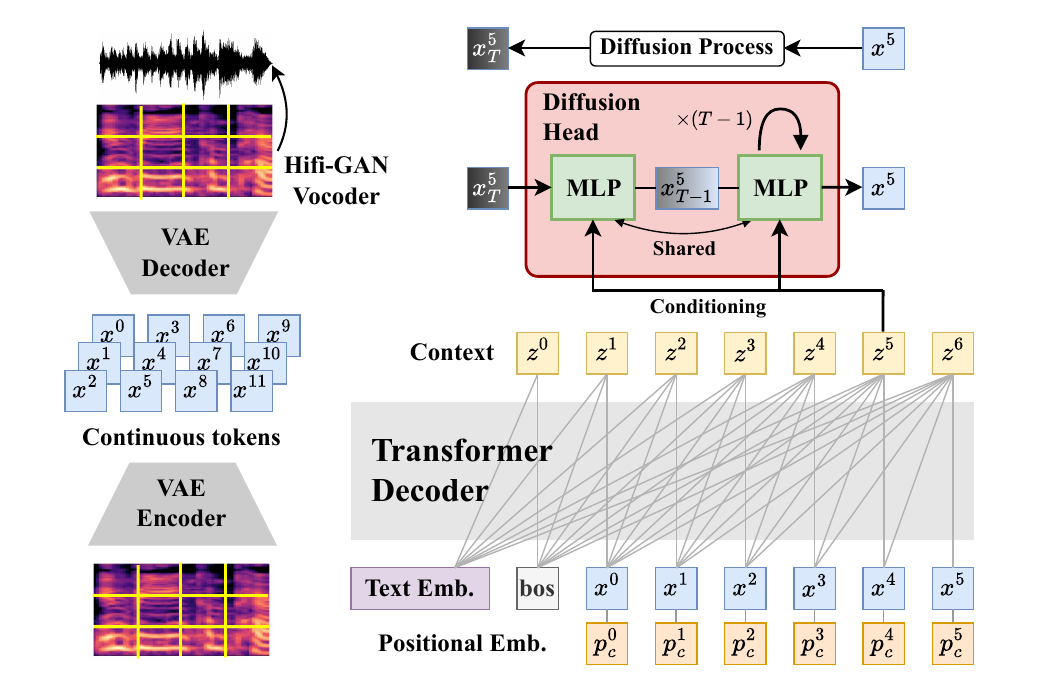}}
\caption{The framework of the continuous-valued audio token proposal. The waveform is transformed into the Mel-spectrogram, subsequently encoded into continuous-valued tokens. The Transformer decoder learns the next token prediction on these tokens via the token-wise diffusion loss with a small MLP diffusion head. This framework is termed \textbf{AudioNTP}.}
\label{image:AudioNTP}
\end{center}
\vskip -0.2in
\end{figure}

% next token prediction is not necessary entangled with discrete token and cross entropy loss
Common belief holds that next-token prediction is tied to a fixed size dictionary and cross-entropy loss.
However, the task by definition is \textit{predicting the next token given the previous tokens}, independent of the exact realization of the \textit{token} and \textit{how to model the next-token distribution}.
Following the paradigm shift in CV~\cite{tschannen2025givt,li2024autoregressive}, we investigate the use of continuous-valued tokens in audio language modeling.
Figure~\ref{image:AudioNTP} illustrates the framework of AudioNTP.

\paragraph{Training.} Given an input text prompt $w$ and a length-$s$ audio waveform $a = \{a^1, ..., a^s\}$, we tokenize it into a sequence of low-dimensional latents\footnote{
\label{footnote:tokenization}
See Appendix~\ref{appendix:audio_tokenizer} for the detailed tokenization pipeline.
}: $x = \{x^1, ..., x^n$\}, where $x^i \in R^h$ and $h \in Z$ represents the latent dimension.
We refer to this length-\(n\), 1-D latent sequence \(x\) as the \textit{continuous-valued tokens}.
We train the language model with maximum likelihood on $x$ over an audio corpus $D = \{ a_j, w_j \}$. The loss function is then $\mathbb{E}_{(a, w) \sim D} \left[ - log \ p(a~ |~ w) \right]$, where

\begin{equation}
p(a~ |~ w) = p(x^1, ..., x^{n}~ |~ w) = \prod^n_{i=1}  p(x^i~ |~ x^1, ..., x^{i-1}, w)
\label{eq:autoreg}
\end{equation}

To learn the next-token distribution $p(x^i~ |~ x^1, ..., x^{i-1}, w)$, our model consists of a Transformer decoder\footnote{
\label{footnote:mar_causal_poor}
Compared to MAR~\cite{li2024autoregressive} which relies on a bidirectional Transformer to achieve competitive results, we stick to the causality to align with the LLM setting.
Their use of continuous-valued tokens on causal LM performs poorly, while our approach achieves results comparable to the bidirectional counterpart with the masked next-token prediction.
} $C_\theta$ and a MLP diffusion head $M_\phi$, where $\theta$ and $\phi$ denote the parameters.
Firstly, $C_\theta$ encodes the input sequence $\{ w, \beta, x^1, ... x^{i-1} \}$ into a sequence of context vectors $z = \{ z^1, ..., z^{i} \}$, where $w$ is placed at the front as the input prompt and $\beta$ is the BOS token: $z^i = C_\theta(w, \beta, x^1, ..., x^{i-1})$.
A small multi-layer perceptron (MLP) $M_\phi$ conditions on $z^i$ and models the next-token distribution $p(x^i~ |~ z^i) = p(x^i~ |~ x^1, ..., x^{i-1}, w)$ with the following diffusion objective~\cite{li2024autoregressive,rombach2022high}.

\begin{equation}
\arg\min_{\theta, \phi}~ \mathbb{E}_{\varepsilon, t} \left[ \left\| \varepsilon - M_\phi(x^i_t,~ z^i,~ t) \right\|^2 \right].
\label{eq:denoise}
\end{equation}

$\varepsilon \in \mathbb{R}^d$ is a Gaussian noise sampled from $\mathcal{N}( \mathbf{0}, \mathbf{I})$.
The diffused token $x^i_t$ is obtained by $x^i_t = \sqrt{\abar_t} x^i + \sqrt{1-\abar_t} \e$ with the noise schedule $\abar_t$~\cite{Ho2020,Nichol2021}.
$t \in \{0, ..., T\}$ is a time step sampled from the noise schedule.
We train $C_\theta$ and $M_\phi$ jointly.
By backpropagating the gradient from the small diffusion head $M_\phi$ to the Transformer decoder $C_\theta$, the decoder represents the next-token distribution into the context vector $z^i$. This facilitates accurate diffusion modeling even with a lightweight $M_\phi$.

\paragraph{Inference.} 

At position $i$, $C_\theta$ infers the context vector $z^i$ by a single pass given the previously sampled tokens $z^i = C_\theta(y, \tilde{x}^1, ..., \tilde{x}^{i-1})$.
Conditioning on $z^i$, $M_\phi$ iteratively de-noises a Gaussian noise into a clean token $\tilde{x}^i$, known as the sampling process.
Sampling is done via a reverse diffusion procedure \cite{Ho2020}:

\begin{equation}
\tilde{x}^i_{t-1} = \frac{1}{\sqrt{\alpha_t}} \left( \tilde{x}^i_t -  \frac{1 - \alpha_t}{\sqrt{1 - \abar_t}}~ M_\phi(\tilde{x}^i_t,~ z^i,~ t)     \right) + \sigma_t \delta.
\label{eq:sampling}
\end{equation}

$\delta$ is sampled from $\mathcal{N}( \mathbf{0}, \mathbf{I})$ and $\sigma_t$ is the noise level at time step $t$.
The decoding de-noising procedure starts from the timestamp $T$ to timestamp $0$.
That is, $\tilde{x}^i_{T} \sim \mathcal{N}( \mathbf{0}, \mathbf{I})$ and $\tilde{x}^i_0 = \tilde{x}^i \sim \ p(x^i~ |~ z^i) = p(x^i~ |~ x^1, ..., x^{i-1}, w)$.
The sampled tokens can then be de-tokenized into the sampled waveform\textsuperscript{\ref{footnote:tokenization}}.

\subsection{AudioMNTP: Masked Next-Token Prediction}
\label{section:AudioMNTP}

\begin{figure}[t]
% \vskip 0.2in
\begin{center}
\centerline{\includegraphics[width=1.1\columnwidth]{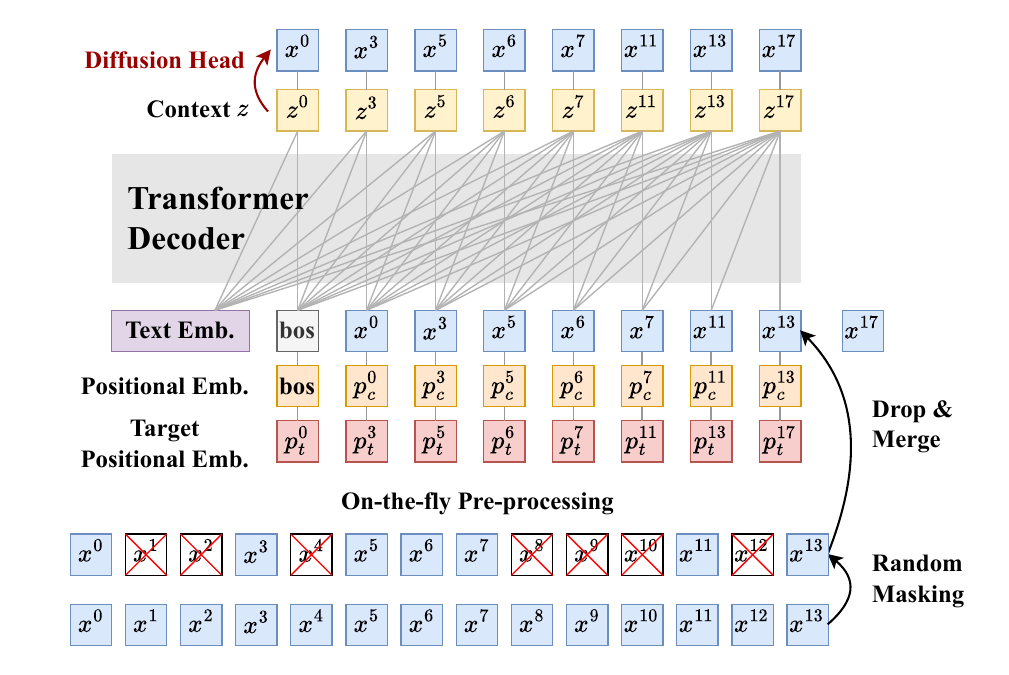}}
\caption{The framework of the masked next-token prediction (MNTP). The continuous-valued tokens are first masked (dropped) to form a shorter sequence. The Transformer decoder then learns the next-token prediction on the dropped sequence with the diffusion loss. This framework is termed \textbf{AudioMNTP}.}
\label{image:AudioMNTP}
\end{center}
\vskip -0.2in
\end{figure}

AudioNTP yields significant improvement over AudioGen.
However, as shown in our results, the generation quality and diversity still lag behind LDMs~\cite{AudioLDM2,tango2}.
Instead of resorting to the bidirectional solution\textsuperscript{\ref{footnote:mar_causal_poor}}, we stick to the causality to maximally follow the LLM framework.
Given the success of MLM in generative models, we seek to migrate the core idea of MLM into the causal next-token prediction.
We hypothesize that the key mechanism behind MLM is \textit{predicting the unseen token given the sparse context}\footnote{
In MAE~\cite{he2022masked} and AudioMAE~\cite{huang2022masked}, over 70\% tokens are masked.
}, essentially independent of the model directionality.
We devise a novel learning task tailored to the causal architectures termed \textbf{masked next-token prediction (MNTP)}.
Figure~\ref{image:AudioMNTP} illustrates the framework.

\paragraph{Training.}

Before feeding the token sequence $x = \{ x^1, ..., x^n \}$ into $C_\theta$, we first apply masking on it.
We denote the mask as $v = \{v^1, ..., v^n$\} where $v^i \in \{0, 1\}$ and $v^i = 0$ means the position $i$ is masked.
The masked sequence is then $\bar{x}_v = x \odot v$.
Recent works show that dropping instead of masking yields similar performances while greatly reduce the training cost\footnote{Due to the drastically reduced sequence length~\cite{huang2022masked,he2022masked,baevski2023efficient}.}.
We drop the tokens to form a shorter sequence $x_v = \{ \bar{x}^i_v \in \bar{x}_v~ |~ \bar{x}^i_v \neq 0 \}$.
We learn the next-token prediction on $x_v$ as the masked prediction\footnote{
Appendix~\ref{appendix:mntp_vs_mlm} reviews the similarity and differences between MLM and MNTP.
}.
Upon a closer look, each position $i$ learns to predict various future positions, depending on the random masking patterns.
We can then view MNTP as a \textit{multi-task learning with input dropout}~\cite{hou2022token,Zaremba2014RecurrentNN,sennrich2016edinburgh}, where \textbf{skip-token prediction} benefits next-token prediction during inference\footnote{
Predicting skip positions have been proven effective for audio representation learning~\cite{oord2018representation,chung2020improved}.
We verify its effectiveness in autoregressive generative modeling. Compared to text, continuous data exhibits high similarity between consecutive tokens, allowing the model to easily exploit local smoothness~\cite{oord2018representation} for next-token prediction.
We avoid this trivial prediction and facilitate a sparse context required by MLM via skip-token prediction.
% Our results verify that skip-token prediction encourages more diversity in next-token prediction inference.
}.
Since the model is unaware of the current masking pattern, it is confusing to naively ask the model predict various future positions indiscriminately.
We introduce the \textbf{target positional embedding} $p_t$ to $C_\theta$, as shown by Figure~\ref{image:AudioMNTP}.
Originally, each token has its own learnable content positional embedding $p_c$.
We learn an additional target positional embeddings representing the position to be predicted and add both embeddings to the audio tokens.
Consequently, masked next-token prediction can be viewed as a generalized form of causal language modeling. 
The model can \textbf{\textit{predict any future timestamp given any subset of past information}}, with \textit{predicting the immediate next token} and \textit{conditioning on all previous tokens} as special cases of the standard next-token prediction\footnote{
Appendix~\ref{appendix:mntp_vs_ntp} illustrates the idea more formally.
}.

\paragraph{Inference.} The inference procedure of AudioMNTP is identical to that of AudioNTP, except for the inclusion of an additional target positional embedding, which is derived by incrementing the content positional embedding by one.

%% file: icml2024/sections/implementation.tex
\section{Implementation}

\paragraph{Continuous-valued Audio Tokenizer.}

We leverage the tokenization pipeline of AudioLDM~\cite{liu2023audioldm}, including a pre-trained VAE and a pre-trained Hifi-GAN~\cite{kong2020hifi} vocoder.
Given an audio, the pipeline extracts the Mel-spectrogram, and the VAE encoder processes it into a sequence of continuous-valued tokens $x = \{ x^1, ..., x^n \}$, which is used for training.
During inference, we sample $\tilde{x} = \{ \tilde{x}^1, ..., \tilde{x}^n \}$ according to equation~\ref{eq:sampling}.
We use the VAE decoder to decode $\tilde{x}$ back to a Mel-spectrogram.
Then, the vocoder is used to synthesize the waveform.
Refer to Appendix~\ref{appendix:audio_tokenizer} for details.

\paragraph{Conditional Audio Language Modeling.}

We mostly follow the implementation in MAR~\cite{li2024autoregressive}, including the training/inference details of the MLP diffusion head, and the architecture design of MLP and the Transformer decoder.
We train an audio version of the bidirectional MAR as the topline (\textbf{AudioMAR}), a causal LM as the baseline (\textbf{AudioNTP}), and our main proposal (\textbf{AudioMNTP}), which is also causal.
For most of the studies, we train AudioMAR, AudioNTP and AudioMNTP with the 193M Base model.
We also train a 462M AudioMNTP Large to study the scaling behavior.
For the text prompt, we use CLAP~\cite{wu2023large} and FLAN-T5~\cite{chung2024scaling} to extract the text embeddings and concatenate them as the input prompt.
Refer to Appendix~\ref{appendix:continuous_language_modeling} for details.

\paragraph{Initialization.}
\label{section:initialization}

By default, AudioMAR, AudioNTP, and AudioMNTP are obtained by fine-tuning the pre-trained image MAR\footnote{Either the Base or Large model in \href{https://github.com/LTH14/mar}{https://github.com/LTH14/mar}.} on audio data. That is, they all start by learning the MAR task on images and diverge by using MAR, NTP, and MNTP task on audio data, respectively. As shown in the subsequent ablation study, this image initialization only slightly boosts performance and is not critical.

\paragraph{Masking Schedule.}
\label{section:mask_schedule}

For each training iteration, we sample a masking ratio from a distribution over $[0, 1]$ defined by the schedule, and randomly drop the tokens according to the ratio.
Specifically, we apply a mixture of normal distribution and truncated normal distribution, where the former emphasizes the high masking ratio for MLM and the latter preserves the long-tailed distribution on the low masking ratio for next-token prediction.
See Appendix~\ref{appendix:mask_schedule} for details.

%% file: icml2024/sections/experiments.tex
\section{Experiments}

\input{icml2024/tables/main}
\input{icml2024/tables/subjective}

\input{icml2024/tables/mntp_ablation}

\paragraph{Training.}

We train our model on AudioCaps (AC)~\cite{kim2019audiocaps} and WavCaps (WC)~\cite{mei2024wavcaps}.
See Appendix~\ref{appendix:data} for details.
We use AdamW~\cite{Loshchilov2017DecoupledWD} optimizer with a fixed learning rate $1.0 \times 10^{-4}$.
We train the Base model with 40 NVIDIA V100 GPUs, and the Large model requires 104.
Our effective batch size is 2048 10-second clips.
We train the Base and the Large model for 1000 epochs, about 2 days and 5 days, respectively.

\paragraph{Evaluation.}

We evaluate our model on the AC evaluation set, following the protocol in AudioLDM and using its evaluation toolkit\footnote{\href{https://github.com/haoheliu/audioldm_eval}{https://github.com/haoheliu/audioldm\_eval}}. For each generated audio and its ground-truth counterpart in AC, we compute several metrics: Fréchet Audio Distance (FAD)~\cite{kilgour19_interspeech}, Fréchet Distance (FD), Kullback--Leibler divergence (KL), Inception Score (IS)~\cite{liu2023audioldm}, and Contrastive Language-Audio Pretraining (CLAP) score~\cite{huang2023make}. See Appendix~\ref{appendix:objective_eval} for details on these metrics.
For subjective evaluation, we rate audio samples on text relevance (REL) and overall quality (OVL) using a 1-5 scale~\cite{tango,liu2023audioldm}; additional details are in Appendix~\ref{appendix:subjective_eval}. Because speech is challenging in TTA---and even the most advanced systems often produce unintelligible speech---we split our evaluation into speech and non-speech categories.

\subsection{Main results}

% We highlight several findings in the main Table~\ref{table:main}.

\paragraph{Continuous-valued tokens with diffusion loss are competitive for audio language modeling.}

Table~\ref{table:main} shows that our 193M baseline AudioNTP Base outperforms the 285M AudioGen Base and 1B UniAudio by a large margin on the FAD and KL metrics.
Specifically, we achieve 20\%, 40\% relative improvements over AudioGen Base on FD and KL scores, respectively.
Both AudioGen and UniAudio are based on the discrete tokens and use much more data, parameters and compute compared to our method.
The results gauge the effectiveness of the continuous-valued tokens and the diffusion loss for audio language modeling\footnote{
We do not train our model with discrete tokens due to computational constraints imposed by their long sequence length. For each 10-second clip, our method uses only 256 tokens, whereas AudioGen uses 5,000 tokens.
% , and UniAudio uses 500 frames—each composed of 3 tokens.
}.

\paragraph{MNTP significantly outperforms next-token prediction.}

Table~\ref{table:main} shows that AudioMNTP Base outperforms our baseline AudioNTP Base by a large margin, with 26\%, 10\%, and 9\% relative improvements on FAD, KL and CLAP scores respectively, demonstrating the effectiveness of MNTP.

\paragraph{Scaling up MNTP reaches SOTA performances on FD and FAD.}

We scale up the model to produce the 462M AudioMNTP Large model.
The scaling boosts the performance across most of the metrics, including the 27\% and 1.5\% relative improvements on FAD and CLAP scores compared to AudioMNTP Base.
Our model is intrinsically less expressive compared to SOTA diffusion models, (1) our model is uni-directional and (2) our model is significantly smaller, i.e. 866M Tango 2.
However, AudioMNTP Large demonstrates the best FD and FAD scores across all the models, and reach the similar KL, IS and CLAP scores compared to the leading diffusion models\footnote{We place 5th, 2nd, and 4th based on the KL, IS, and CLAP scores, respectively. Most of the scores are close. Tango 2’s especially high CLAP score is attributed to the additional preference dataset Audio-Alpaca~\cite{tango2}.}.
% \footnote{Our results further demonstrate that relying on the entire Transformer to predict the diffusion noise, as done in LDMs, is not necessary. Instead, using a small MLP as the diffusion head, conditioned by a strong context vector $z$, is sufficient.}.
We believe performance can be improved by further scaling up the model and data, and we leave this for future work due to computational constraints.

\subsection{Subjective evaluation}

We compare with the best existing decoder-only solution, AudioGen, and two leading diffusion models, AudioLDM 2 and Tango 2.
Table~\ref{table:subjective} shows that AudioMNTP is significantly better than AudioGen in both speech and non-speech categories, and approaching the performance of Tango 2.
In both categories, our REL scores lag behind those of Tango 2; however, Tango 2 is a bi-directional model and utilizes the preference optimization dataset to enhance text-audio alignment.
Interesting, we find that our method is especially good at generating authentic speech, indicated by the highest OVL score in the speech category, potentially due to the incorporation of MLM, the de facto standard in speech pre-training~\cite{liugenerative,baevski2020wav2vec}\footnote{MLM is effective in both discriminative~\cite{baevski2020wav2vec} and generative~\cite{liugenerative} speech pre-training.}.

\subsection{Ablation studies}
\label{section:mntp_ablation}

We ablate the components of AudioMNTP in Table~\ref{table:ablate_mntp_results} and visualize them in the Appendix~\ref{appendix:ablate_mntp}.
Note that different masking strategies, such as masking or dropping, favor different masking schedules.
We explored various schedules and report the best result.

% \begin{figure}[t]
% \begin{center}
% \centerline{\includegraphics[width=\columnwidth]{icml2024/images/mntp-ablation.pdf}}
% \caption{Visualizing the differences between the ablation variants. The subfigure IDs match the IDs in Table~\ref{table:ablate_mntp_results}.}
% \label{image:mntp_ablation}
% \end{center}
% \vskip -0.2in
% \end{figure}

\paragraph{Combining MLM and NTP.}

Table~\ref{table:ablate_mntp_results} (B) naively combines MLM with NTP by using the bidirectional AudioMAR as the initialization and fine-tuning with next-token prediction on audio data.
That is, Table~\ref{table:ablate_mntp_results} (A) uses the image MAR as the initialization, but Table~\ref{table:ablate_mntp_results} (B) uses AudioMAR as the initialization, which additionally learns the MAR task on audio data.
Table~\ref{table:ablate_mntp_results} (B) shows that MLM on audio data improves performance, but only slightly.

\paragraph{Migrate MLM into NTP: masking.}

The bidirectional MLM is essentially not designed for the causal decoding.
We investigate incorporating input masking into next-token prediction.
Table~\ref{table:ablate_mntp_results} (C) follows MAR to apply zero masking, but uses the causal decoder for next-token prediction.
Table~\ref{table:ablate_mntp_results} (D) tries the Gaussian noises as the masking, since the numerical value space is more aligned to that of the clean tokens, which are sampled from the same Gaussian by the diffusion head.
Appendix~\ref{appendix:ablate_mntp} Figure~\ref{image:mntp_ablation} (C) illustrates the idea.
Table~\ref{table:ablate_mntp_results} (E) tries dropping the tokens as a type of masking, which significantly reduces the compute cost, as shown by Appendix~\ref{appendix:ablate_mntp} Figure~\ref{image:mntp_ablation} (E).
We predict the next tokens directly on top of the visible tokens instead of the masked tokens, and the masked tokens are completely removed.
Given that our masking ratio is typically above 70\%~\cite{li2024autoregressive,huang2022masked}, dropping only requires less than half of the original computation during training.
According to Table~\ref{table:ablate_mntp_results} (C), (D), and (E), incorporating input masking to next-token prediction improves the performance on most metrics compared to Table~\ref{table:ablate_mntp_results} (A), including FD, FAD and CLAP, except for IS.
We proceed with the dropping scenario due to its computational efficiency. However, our proposal is independent of the exact realization of the masking.
Note that other masking strategies, such as zero masking, do not work well with our final masking schedule, as indicated in Table~\ref{table:ablate_mntp_results} (F).
In contrast, dropping tokens remains effective with different masking schedules in subsequent studies.

\paragraph{Migrate MLM into NTP: masked prediction.}

We study the effectiveness of skip-token prediction, our form of masked prediction, which we hypothesize to benefit next-token prediction.
Table~\ref{table:ablate_mntp_results} (G), AudioMNTP, justifies our hypothesis and outperforms all the rows on all metrics.
We ablate our mixture of distributions masking schedule to verify the robustness of MNTP.
We replace the schedule by the one used in MAR\footnote{We find that MAR's truncated normal distribution on $[0.7, 1.0]$ works poorly, but simply shifting it to the left to $[0.55, 0.85]$ works reasonably well.}.
Table~\ref{table:ablate_mntp_results} (H) shows that our MNTP task, even with the masking schedule of MAR, remains competitive compared to Table~\ref{table:ablate_mntp_results} (A) and Table~\ref{table:ablate_mntp_results} (B), only slightly underperforming Table~\ref{table:ablate_mntp_results} (G). This demonstrates MNTP's robustness to the masking schedule.
Next, we ablate the target positional embedding used for skip-token prediction.
Table~\ref{table:ablate_mntp_results} (I) is significantly worse than Table~\ref{table:ablate_mntp_results} (G) on all metrics, suggesting that different future positions conflict each other.
Table~\ref{table:ablate_mntp_results} (I) is even worse than Table~\ref{table:ablate_mntp_results} (E), which simply predicts the immediate next token.
The results demonstrate the importance of the target positional embedding.

\paragraph{Initialization.}
\label{section:init_ablation}

We replace the image MAR initialization by random initialization for AudioMNTP.
Table~\ref{table:ablate_mntp_results} (J) underperforms Table~\ref{table:ablate_mntp_results} (G) on all metrics, but slightly, suggesting that the image MAR initialization is helpful but not necessary for MNTP to work.

\input{icml2024/tables/mlp}
\input{icml2024/tables/text_embedding}
\input{icml2024/tables/causal_decode}

\paragraph{The size of the MLP diffusion head.}

Table~\ref{table:mlp_size} shows that the size of the diffusion head matters.
Increasing the head from 1 layer to 3 layers significantly improve the performances on most metrics.
Further increasing to 6 layers significantly decrease the FAD score.
However, with the larger MLP size, the inference time increases.
Our results highlight that the size of the diffusion head is an important hyperparameter for balancing performance and efficiency during trade-offs.

\paragraph{The ensemble text embedding.}

Table~\ref{table:text_emb} shows that concatenating CLAP and FLAN-T5 yields the best performance, as both embeddings provide complementary information to the model, and neither individual embedding can match the performance of the joint embeddings.

\subsection{MAR vs. MNTP vs. Next-Token Prediction}

% We compare three methods in Table~\ref{table:causal_decode}.

\paragraph{MNTP is approaching MAR.}

First, Table~\ref{table:causal_decode}A indicates that AudioMAR yields the best results on most metrics except for FD.
This is expected, as it is a bidirectional model with the full context.
However, compared to AudioNTP, AudioMNTP significantly bridges the gap and achieves performance comparable to that of AudioMAR on IS and CLAP score.
AudioMNTP even surpasses AudioMAR on FD and KL, suggesting that MNTP is a competitive task for learning causal language modeling on continuous data.

\paragraph{The effectiveness of MNTP in causal decoding.}

MAR is competitive, but its use of future context does not support streaming in the LLM framework well.
We verify this hypothesis by comparing AudioMAR, AudioNTP, and AudioMNTP under causal decoding.
AudioNTP and AudioMNTP are intrinsically developed for causal decoding, while AudioMAR can perform causal decoding by unmasking from left to right.
Causal decoding degrades the performance of AudioMAR when comparing Table~\ref{table:causal_decode} (A) and (C).
Interesting, by increasing the decoding steps as shown by Table~\ref{table:causal_decode} (D), AudioMAR works much better, and even surpasses the naive AudioNTP in Table~\ref{table:causal_decode} (E).
However, Table~\ref{table:causal_decode} (F) indicates that AudioMNTP is the best in most metrics in the causal decoding scenario\footnote{
We verified the idea of simply applying MLM to a Transformer decoder in Table~\ref{table:ablate_mntp_results} (C) and (D), which are worse than AudioMNTP.
}.

\subsection{Inference Speed vs. Generation Quality}

\begin{figure}[t]
% \vskip 0.2in
\begin{center}
\centerline{\includegraphics[width=1\columnwidth]{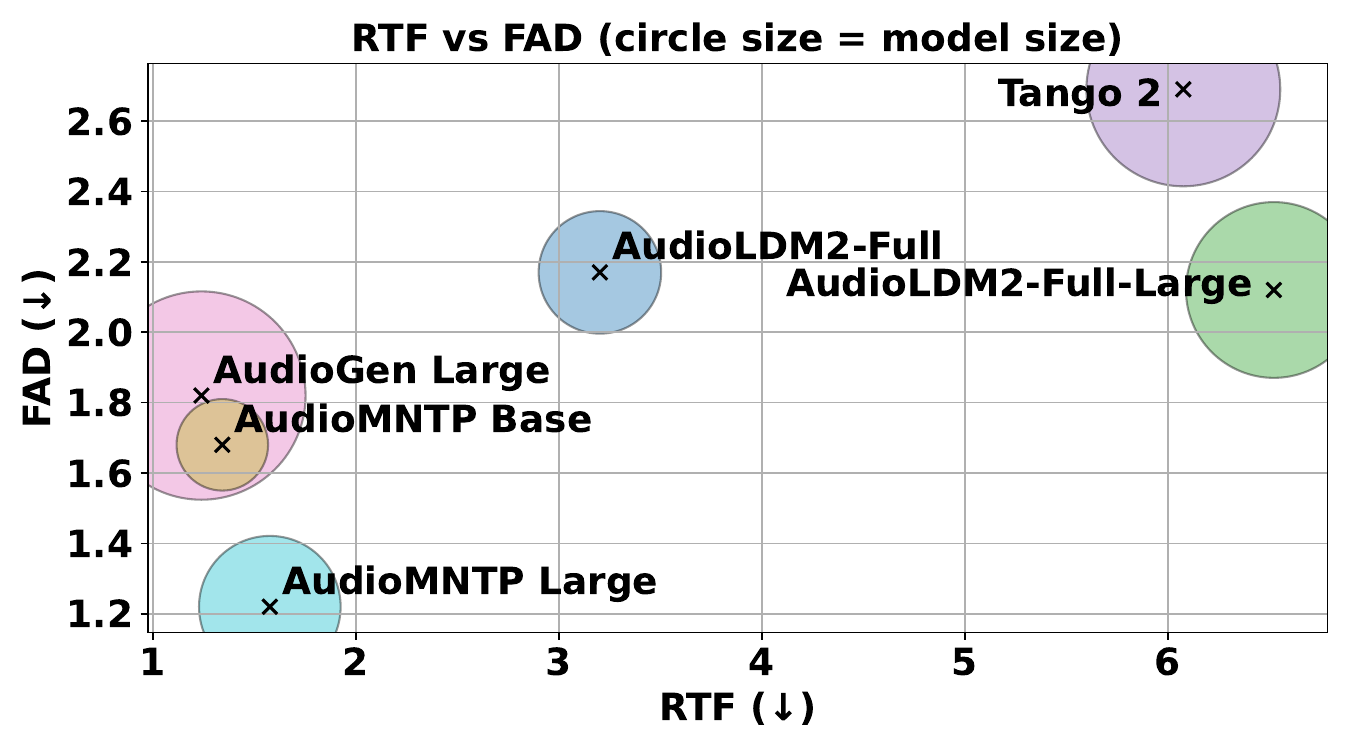}}
\caption{\textbf{Comparing the speed and quality tradeoff.} The RTF is real-time factor, calculated by $\frac{\textbf{processing duration}}{\textbf{clip duration}}$. Our AudioMNTP Large reach near real-time streamable generation with a single NVIDIA A100 GPU. Both AudioMNTP Base and Large achieve the lowest FAD with the smallest model size.}
\label{image:fad_vs_rtf}
\end{center}
\vskip -0.2in
\end{figure}

We compare the speed and quality tradeoff in Figure~\ref{image:fad_vs_rtf}, using AudioGen Large\footnote{AudioGen Base was not released.} for discrete LMs, AudioLDM 2 Large and Tango 2 for diffusion models, and our AudioNTP and AudioMNTP Base/Large.
The real-time-factor (RTF) is calculated by $\frac{\textbf{processing duration}}{\textbf{clip duration}}$. Both durations are measured in seconds. Processing duration is obtained by a single NVIDIA A100 GPU with batch size 1.
Figure~\ref{image:fad_vs_rtf} shows that AudioMNTP Base and Large achieve the best FAD while being streamable and near real-time generation. The real-time generation is easily achieve by switching to NVIDIA H100 GPU.
Diffusion models are much slower than LMs and continuous-valued LM with MNTP is much better than discrete-valued LM.

%% file: icml2024/tables/main.tex
\begin{table*}[t]
\caption{\textbf{Main results.} FD, FAD, KL, IS, and CLAP metrics on the AudioCaps evaluation set. The AS, AC, and WC stand for AudioSet, AudioCaps, and WavCaps, respectively. \textbf{Bold}: the best performance. \underline{Underline}: the second best performance. \textsuperscript{*}Re-inference with the public checkpoint. The detailed datasets used by each model are listed in the Appendix~\ref{appendix:data}.}
\label{table:main}
\vskip 0.05in
\centering
\begin{sc}
\resizebox{\linewidth}{!}{
    \begin{tabular}{llcc|ccccc}
    \toprule

    Token & Method & Datasets & \#Params & FD $\downarrow$ & FAD $\downarrow$ & KL $\downarrow$ & IS $\uparrow$ & CLAP $\uparrow$ \\

    \midrule

    \textbf{Bi-directional} \\
    \midrule

    \multirow{2}{*}{Discrete} & MAGNET-Small~\cite{magnet} & {AS + AC + 8 others} & 300M & 23.02 & 3.22 & 1.42 & 9.72 & 0.287 \\

     {} & MAGNET-Large~\cite{magnet} & {AS + AC + 8 others} & 1.5B & 26.19 & 2.36 & 1.64 & 9.10 & 0.253 \\

    \cmidrule(lr){1-2}

    \multirow{9}{*}{Continuous} & Tango~\cite{tango} & {AC} & 866M & 24.52 & 1.59 & 1.37 & 7.70 & 0.313 \\

    {} & Tango-Full-FT~\cite{tango} & {AS + AC + WC + 5 others} & 866M & 18.93 & 2.19 & 1.12 & 8.80 & 0.340 \\

    {} & Tango-AF\&AC-FT~\cite{kong2024improving} & {AC + 1 other} & 866M & 21.84 & 2.35 & 1.32 & 9.59 & 0.343 \\

    {} & Tango 2~\cite{tango2} & {AS + AC + WC + 6 others} & 866M & 20.66 & 2.69 & 1.12 & 9.09 & \textbf{0.375} \\

    {} & Make-an-Audio 2~\cite{make-an-audio2} & {AS + AC + WC + 11 others} & 937M & 16.23 & 2.03 & 1.29 & \textbf{9.95} & \underline{0.345} \\

    % {} & EzAudio-L~\cite{EZAudio} & {AS + AC + 3 others} & 596M & 15.59 & 2.25 & 1.38 & \underline{11.35} & 0.391 \\

    % {} & EzAudio-XL~\cite{EZAudio} & {AS + AC + 3 others} & 874M & 14.98 & 3.01 & 1.29 & \textbf{11.38} & 0.387 \\

    {} & AudioLDM2-AC~\cite{AudioLDM2} & {AC} & 346M & - & 1.67 & \underline{1.10} & - & - \\

    {} & AudioLDM2-AC-Large~\cite{AudioLDM2} & {AC} & 712M & - & \underline{1.42} & \textbf{0.98} & - & - \\

    {} & AudioLDM2-Full~\cite{AudioLDM2} & {AS + AC + WC + 2 others} & 346M & - & 1.78 & 1.60 & - & - \\

    {} & \ \ - re-inference\textsuperscript{*} & {} & 346M & 32.14 & 2.17 & 1.62 & 6.92 & 0.273 \\

    {} & AudioLDM2-Full-Large~\cite{AudioLDM2}& {AS + AC + WC + 2 others} & 712M & - & 1.86 & 1.64 & -	& - \\

    {} & \ \ - re-inference\textsuperscript{*} & {} & 712M & 33.18 & 2.12 & 1.54 & 8.29 & 0.281 \\

    \midrule
    \midrule

    \textbf{Uni-directional}\\
    \midrule

    \multirow{3}{*}{Discrete} & AudioGen Base~\cite{AudioGen} & {AS + AC + 8 others} & 285M & - & 2.84 & 2.14 & - & - \\

    {} & AudioGen Large~\cite{AudioGen} & {AS + AC + 8 others} & 1B & - & 1.82 & 1.69 & -	& - \\

    {} & UniAudio~\cite{uniaudio} & {AS + AC + WC + 10 others} & 1B & - & 3.12 & 2.60 & -	& - \\

    \cmidrule(lr){1-2}

    \multirow{4}{*}{Continuous}  & \textbf{Ours} \\

    \cmidrule(lr){2-2}

    {} & \textbf{AudioNTP Base} & {AC + WC} & 193M & 18.52 & 2.28 & 1.29 & 9.42 & 0.308 \\

    {} & \textbf{AudioMNTP Base} & {AC + WC} & 193M & \underline{14.81} & 1.68 & 1.16 & 9.67 & 0.336 \\

    {} & \textbf{AudioMNTP Large} & {AC + WC} & 462M & \textbf{14.30} & \textbf{1.22} & 1.17 & \underline{9.81} & 0.341 \\

    \bottomrule
    \end{tabular}
}
\end{sc}
% \end{small}
\vskip -0.1in
\end{table*}

%% file: icml2024/tables/subjective.tex
\begin{table}[t]
\caption{Human Evaluation Results. We use AudioMNTP Large to compare to other models.}
\label{table:subjective}
\vskip 0.05in
\centering
\begin{sc}
\resizebox{1.01\linewidth}{!}{
    \begin{tabular}{lcc|cc}
    \toprule
    & \multicolumn{2}{c}{\textbf{Non-Speech}} & \multicolumn{2}{c}{\textbf{Speech}}
    \\
    \cmidrule(lr){2-3}
    \cmidrule(lr){4-5}
     Method & REL $\uparrow$ & OVL $\uparrow$ & REL $\uparrow$ & OVL $\uparrow$ \\
    \midrule

    Reference & 4.06 $\pm$ 1.09 & 3.87 $\pm$ 1.11 & 4.47 $\pm$ 0.83 & 4.61 $\pm$ 0.95  \\
    \midrule
    \multicolumn{5}{l}{\textbf{bi-directional}} \\
    \cmidrule(lr){1-1}
    AudioLDM 2 & 3.10 $\pm$ 1.29 & 3.32 $\pm$ 1.17 & 3.17 $\pm$ 1.14 & 3.01 $\pm$ 0.91  \\
    Tango 2 & 3.95 $\pm$ 0.97 & 3.80 $\pm$ 0.99 & 4.20 $\pm$ 0.79 & 3.15 $\pm$ 0.94 \\
    \midrule
    \midrule
    \multicolumn{5}{l}{\textbf{uni-directional}} \\
    \cmidrule(lr){1-1}
    AudioGen & 3.05 $\pm$ 1.09 & 3.02 $\pm$ 1.17 & 3.33 $\pm$ 0.97 & 2.56 $\pm$ 1.13\\
    \textbf{AudioMNTP} & 3.79 $\pm$ 1.02 & 3.46 $\pm$ 1.10 & 3.91 $\pm$ 0.94 & 3.69 $\pm$ 0.88 \\
    \bottomrule
    \end{tabular}
}
\end{sc}
% \end{small}
\vskip -0.1in
\end{table}

%% file: icml2024/tables/mntp_ablation.tex
\newcommand{\checking}{\color{Green} \ding{51}}
\newcommand{\crossing}{\color{Red} \ding{55}}

\begin{table*}[t]
\caption{Ablate the components of MNTP with the Base configuration. INIT. means initialization; POS. EMB. means positional embedding.
By default, the masking schedule is the mixture of the normal and truncated normal distribution, as described in Section~\ref{section:mask_schedule}.
\textsuperscript{*}indicates the masking schedule used in MAR~\cite{li2024autoregressive}, a truncated normal distribution over $[0.7, 1]$.
\textsuperscript{\textdagger}indicates the fixed masking ratio at $0.7$.
\textsuperscript{\textdaggerdbl}indicates a masking schedule of uniform distribution over $[0, 1]$.
}
\label{table:ablate_mntp_results}
\vskip 0.05in
\centering
\begin{sc}
\resizebox{\linewidth}{!}{
    \begin{tabular}{l|cc|ccc|cccc|ccccc}
    \toprule

    % \multicolumn{3}{c}{\textbf{misc.}} & \multicolumn{3}{c}{\textbf{How to mask}} & \multicolumn{4}{c}{\textbf{what to predict}} & \multicolumn{5}{c}{\textbf{objective metrics}} \\

    % \midrule

    ID & Causal & MAR init. & \makecell{Zero \\ mask} & \makecell{Gaussian \\ mask} & \makecell{drop \\ token} & \makecell{predict \\ next} & \makecell{predict \\ skip} & \makecell{target \\ pos. emb.} & \makecell{predict \\ masked} & FD $\downarrow$ & FAD $\downarrow$ & KL $\downarrow$ & IS $\uparrow$ & CLAP $\uparrow$ \\

    \midrule

    \multirow{2}{*}{\textbf{(a)}} & \multicolumn{14}{c}{Baseline - AudioNTP} \\

    % \cmidrule(lr){2-15}

    {} & {\checking} & {\checking} & {\crossing} & {\crossing} & {\crossing} & {\checking} & {\crossing} & {\crossing} & {\crossing} & 18.52 & 2.28 & 1.29 & 9.42 & 0.308 \\

    \midrule
    \midrule

    {\textbf{(b)}} & {{\crossing} \ding{221} {\checking}} & {\checking}  & {\checking}\textsuperscript{*} & {\crossing} & {\crossing} & {\checking} & {\crossing} & {\crossing} & {\checking} & 17.15 & 2.13 & 1.25 & 9.45 & 0.321 \\

    % \midrule

    {\textbf{(c)}} & {\checking} & {\checking} & {\checking}\textsuperscript{\textdagger} & {\crossing} & {\crossing} & {\checking} & {\crossing} & {\crossing} & {\crossing} & 16.78 & 1.97 & 1.28 & 9.33 & 0.333  \\

    {\textbf{(d)}} & {\checking} & {\checking} & {\crossing} & {\checking}\textsuperscript{\textdaggerdbl} & {\crossing} & {\checking} & {\crossing} & {\crossing} & {\crossing} & 15.15 & 1.82 & 1.18 & 9.22 & 0.324 \\

    {\textbf{(e)}} & {\checking} & {\checking} & {\crossing} & {\crossing} & {\checking} & {\checking} & {\crossing} & {\crossing} & {\crossing} & 16.62 & 1.77 & 1.32 & 9.25 & 0.315 \\

    {\textbf{(f)}} & {\checking} & {\checking} & {\checking} & {\crossing} & {\crossing} & {\checking} & {\crossing} & {\crossing} & {\crossing} & 24.45 & 4.49 & 1.68 & 5.87 & 0.229 \\

    \midrule
    \midrule

    \multirow{2}{*}{\textbf{(g)}} & \multicolumn{14}{c}{AudioMNTP} \\

    % \cmidrule(lr){2-15}

    {} & {\checking} & {\checking} & {\crossing} & {\crossing} & {\checking} & {\checking} & {\checking} & {\checking} & {\crossing} & \textbf{14.81} & \textbf{1.68} & \textbf{1.16} & \textbf{9.67} & \textbf{0.336} \\

    \midrule
    \midrule

    {\textbf{(h)}} & {\checking} & {\checking} & {\crossing} & {\crossing} & {\checking}\textsuperscript{*} & {\checking} & {\checking} & {\checking} & {\crossing} & 15.55 & 1.89 & \textbf{1.16} & 9.56 & 0.327 \\

    {\textbf{(i)}} & {\checking} & {\checking} & {\crossing} & {\crossing} & {\checking} & {\checking} & {\checking} & {\crossing} & {\crossing} & 20.82 & 3.12 & 1.55 & 8.13 & 0.301 \\

    {\textbf{(j)}} & {\checking} & {\crossing} & {\crossing} & {\crossing} & {\checking} & {\checking} & {\checking} & {\checking} & {\crossing} & 14.85 & 1.70 & 1.17 & 9.62 & 0.335 \\

    \midrule
    \midrule

    \multirow{2}{*}{\textbf{(k)}} & \multicolumn{14}{c}{Topline - AudioMAR} \\

    % \cmidrule(lr){2-15}

    {} & {\crossing} & {\checking} & {\checking}\textsuperscript{*} & {\crossing} & {\crossing} & {\crossing} & {\crossing} & {\crossing} & {\checking} & 14.86 & 1.35 & 1.17 & 9.75 & 0.346 \\

    \bottomrule
    \end{tabular}
}
\end{sc}
% \end{small}
\vskip -0.1in
\end{table*}

%% file: icml2024/tables/mlp.tex
\begin{table}[t]
\caption{\textbf{Ablating different MLP diffusion head sizes} with the AudioMNTP Base configuration. The dimension of each MLP layer is 1024. $L$ denotes the layers of the MLP. The relative size of the heads in the whole model is 5.4\%, 10.7\%, and 17.4\% for $L=1, 3, 6$, respectively. $S$ denotes the inference speed, measured by seconds per 10-second audio with batch size 40 on an NVIDIA V100 GPU.}
\label{table:mlp_size}
\vskip 0.05in
\centering
\begin{sc}
\resizebox{\linewidth}{!}{
    \begin{tabular}{cc|ccccc|c}
    \toprule
    \multicolumn{2}{c}{\textbf{MLP}} & \multirow{2}{*}{FD $\downarrow$} & \multirow{2}{*}{FAD $\downarrow$} & \multirow{2}{*}{KL $\downarrow$} & \multirow{2}{*}{IS $\uparrow$} & \multirow{2}{*}{CLAP $\uparrow$} & \multirow{2}{*}{$S$} \\
    $L$ & Size & {} & {} & {} & {} & {} & {} \\
    \midrule
    1 & 9.84M & 16.97 & 1.68 & 1.20 & 8.87 & 0.321 & \textbf{3.22} \\
    3 & 20.34M & 14.81 & 1.68 & 1.16 & 9.67 & 0.336 & 3.93 \\
    6 & 36.09M & \textbf{14.74} & \textbf{1.21} & \textbf{1.15} & \textbf{9.78} & \textbf{0.337} & 4.84 \\
    \bottomrule
    \end{tabular}
}
\end{sc}
% \end{small}
\vskip -0.1in
\end{table}

%% file: icml2024/tables/text_embedding.tex
\begin{table}[t]
\caption{Comparing the text embedding with the Base configuration. EMB. denotes embedding.}
\label{table:text_emb}
\vskip 0.05in
\centering
\begin{sc}
\resizebox{\linewidth}{!}{
    \begin{tabular}{l|ccccc}
    \toprule
    Text Emb. & FD $\downarrow$ & FAD $\downarrow$ & KL $\downarrow$ & IS $\uparrow$ & CLAP $\uparrow$ \\

    \midrule

    CLAP + FLAN-T5 & \textbf{14.82} & \textbf{1.68} & \textbf{1.16} & \textbf{9.67} & \textbf{0.336} \\
    CLAP & 15.94 & 1.75 & 1.29 & 8.36 & 0.297 \\
    FLAN-T5 & 16.43 & 1.75 & 1.37 & 8.22 & 0.285 \\

    \bottomrule
    \end{tabular}
}
\end{sc}
% \end{small}
\vskip -0.1in
\end{table}

%% file: icml2024/tables/causal_decode.tex
\begin{table}[t]
\caption{\textbf{Comparing MAR and MNTP on different decoding scenarios.} The random-order decoding is the default decoding method used in MAR. The \textbf{steps} means the number of decoding steps. 64 is the default decoding step of MAR. 256 is the total sequence length of the 10-second audio. AudioMAR supports parallel decoding so it can reduce the decoding steps, denoted by \textsuperscript{\dag}.
The \textbf{bold} denotes the best performance with the causal decoding.
The \underline{underline} denotes the globally best performance.
}
\label{table:causal_decode}
\vskip 0.05in
\centering
\begin{sc}
\resizebox{\linewidth}{!}{
    \begin{tabular}{l|lcccccc}
    \toprule

    ID & Model & Steps & FD $\downarrow$ & FAD $\downarrow$ & KL $\downarrow$ & IS $\uparrow$ & CLAP $\uparrow$ \\

    \midrule

    \multicolumn{8}{c}{\textbf{random decoding}} \\
    \midrule

    (A) & \multirow{2}{*}{AudioMAR} & 64\textsuperscript{\dag} & 14.86 & \underline{1.35} & 1.17 & \underline{9.75} & \underline{0.35} \\
    (B) & {} & 256 & 15.02 & 1.57 & \underline{1.13} & 9.47 & 0.347 \\

    \midrule
    \midrule
    \multicolumn{8}{c}{\textbf{causal decoding}} \\
    \midrule

    (C) & \multirow{2}{*}{AudioMAR} & 64\textsuperscript{\dag} & 21.43 & 2.17 & 1.17 & 7.31 & 0.287 \\
    (D) & {} & 256 & 16.32 & 1.84 & \textbf{1.14} & 9.09 & 0.317 \\

    (E) & AudioNTP & 256 & 18.52 & 2.28 & 1.29 & 9.42 & 0.308 \\

    (F) & AudioMNTP & 256 & \underline{\textbf{14.82}} & \textbf{1.68} & 1.16 & \textbf{9.67} & \textbf{0.336} \\

    \bottomrule
    \end{tabular}
}
\end{sc}
% \end{small}
\vskip -0.1in
\end{table}

%% file: icml2024/sections/conclusion.tex
\section{Conclusion}

We propose \textbf{AudioNTP} and \textbf{AudioMNTP}, a high-performing framework for audio language modeling.
Our results demonstrate that \textbf{continuous-valued tokens} are competitive for audio language modeling, yielding 20\% improvements over the discrete token-based audio language models.
Our \textbf{masked next-token prediction (MNTP)} further validates the benefits of applying masked language modeling concept to causal language modeling on continuous data, achieving another 20\% improvements over next-token prediction while preserving the causality during inference.
By combining continuous-valued tokens with MNTP, our approach achieves SOTA audio generation quality, rivaling the leading latent diffusion models and reviving the language modeling approach for sound generation.
Moreover, our models are streamable, significantly smaller and faster than most existing solutions, showcasing the efficiency and effectiveness of the proposed learning techniques. These findings point toward a promising direction for directly generating high-fidelity audio with the streamable and scalable large language models.

%% file: icml2024/sections/appendix.tex
\section{Data}
\label{appendix:data}

We compare the datasets used by our method to the existing systems. We only use AudioCaps~\cite{kim2019audiocaps} and WavCaps~\cite{mei2024wavcaps}, highlighting the data efficiency of our methods.
Audios longer than 10 seconds are randomly cropped into 10 second. That is, the number of the text-audio pairs are the same after the pre-processing.
To speedup the training, we pre-extract the text embedding and the continuous-valued audio tokens.

\input{icml2024/tables/datasets}

\section{Masking Schedules}
\label{appendix:mask_schedule}

\subsection{Defining masking schedules.}
\label{appendix:mask_schedule_define}

\begin{figure}[ht]
% \vskip 0.2in
\begin{center}
\centerline{\includegraphics[width=0.95\columnwidth]{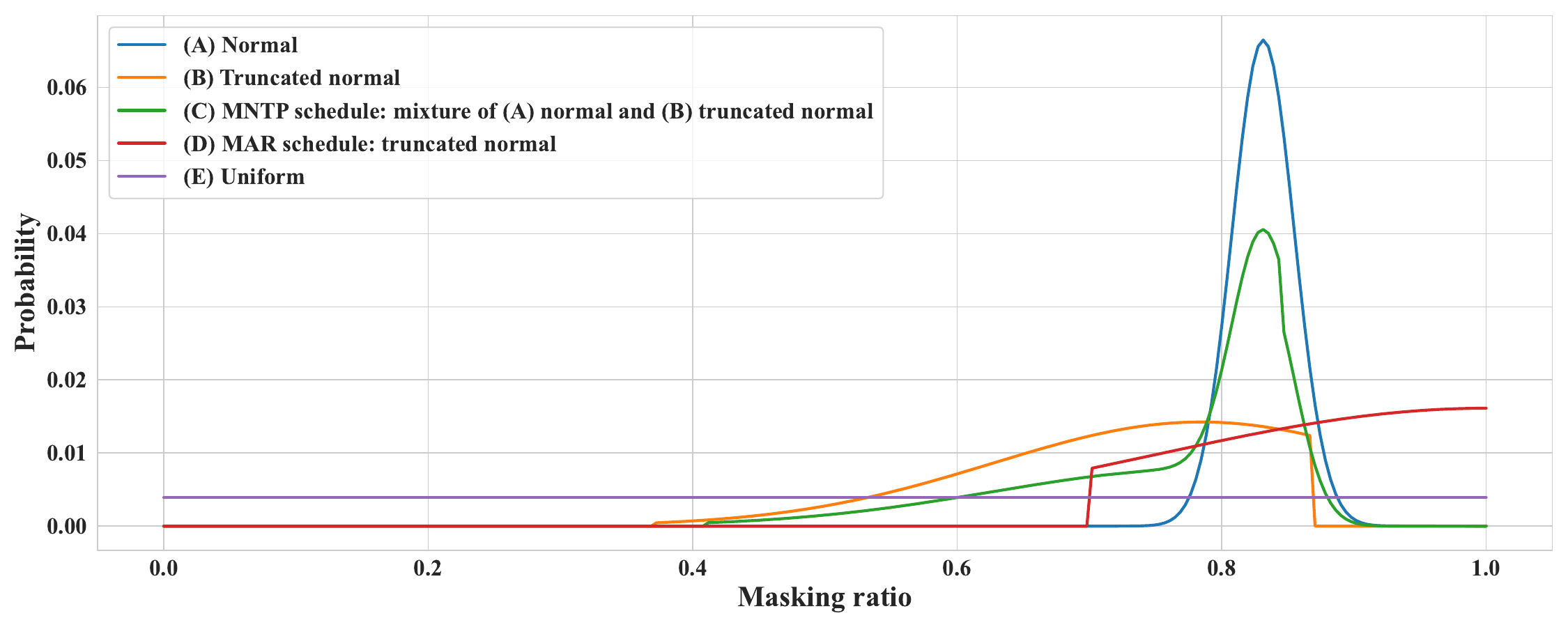}}
\caption{Visualizing the masking schedules. We first sample a masking ratio from the schedule, a probability distribution over $[0, 1]$, and then sample the masking positions in the token sequence based on the ratio. In (C), we average (A) and (B) with equal weights.}
\label{image:masking_schedule}
\end{center}
% \vskip -0.2in
\end{figure}

We explored various masking schedules, including the truncated normal distribution used in MAR, the fixed masking ratio, uniform distribution, and our mixture of normal and truncated normal distribution. Different \textit{masking strategies}, including zero masking, Gaussian masking, and dropping, work best under different \textit{masking schedules}, as suggested by Table~\ref{table:ablate_mntp_results}.
By default, a masking schedule represents a distribution over $[0, 1]$, where we sample a masking ratio $r \sim [0, 1]$ for each training iteration.
Then, given a length-$n$ sequence of continuous-valued tokens, we mask (drop) $n \times r$ tokens.
For the dropping masking strategy, the remaining sequence length is $n \times (1 - r)$.
In Figure~\ref{image:masking_schedule}, we visualize a few representative schedules we find useful in this study.

\subsection{Ablating the MNTP schedule.}
\label{appendix:mask_schedule_result}

We examine the components of the MNTP masking schedule, specifically the normal distribution and the truncated normal distribution. Table~\ref{table:mntp_schedule} shows that the normal distribution (A) plays a more critical role than the truncated normal distribution (B), underscoring the importance of a high masking ratio during MLM. However, the normal distribution alone results in a poor IS score.
We hypothesize that this issue arises from a training/testing mismatch. During next-token prediction, all previous tokens are presented—a scenario that is less frequently encountered during training if the focus is solely on a high masking ratio.
To address this, we incorporate a long-tailed distribution (B) into the normal distribution by averaging them, resulting in (C), our final masking schedule.
As shown in Table~\ref{table:mntp_schedule}, (C) outperforms (D), the MAR default masking schedule, across all metrics, demonstrating the effectiveness of our approach.

\input{icml2024/tables/mask_schedule}

\section{Classifier-free guidance (CFG) and sampling temperature}
\label{appendix:cfg}

\paragraph{Classifier-free guidance.}

Classifier-free guidance is a method for trading off diversity in favor of fidelity.
We reuse the notations in Section~\ref{section:AudioNTP}.
During training, the input text prompt $w$ is replaced with a fake latent $w_f$ with a probability of 10\%.
The fake latent has the same length as the text prompt and is learned jointly with the whole model.
During testing, we run two inferences for each text prompt: conditional generation and unconditional generation.
Specifically, at each decoding position $i$, we get two conditioning vectors: $z_c^i = C_\theta(w, \beta, x^1, ..., x^{i-1})$ and $z_u^i = C_\theta(w_f, \beta, x^1, ..., x^{i-1})$.
At each diffusion sampling step $t$, we then induce two noise predictions with the diffusion head: $\e_c = M_\phi(\tilde{x}^i_t,~ z_c^i,~ t)$ and $\e_u = M_\phi(\tilde{x}^i_t,~ z_u^i,~ t)$.
The classifier-free guidance is realized by linear interpolating two predicted noises.

\begin{equation}
\e = \e_c + \omega_i \cdot  (\e_c - \e_u)
\label{eq:cfg}
\end{equation}

where $\omega_i \in R$ is the guidance scale when decoding the position $i$.
Intuitively, the guidance scale represents the degree of \textit{moving away} from the unconditional generation.
With higher $\omega_i$, the generation becomes more aligned with the given text prompt, albeit with reduced diversity.
We do not fix a guidance scale $\omega_i$ for all positions.
We adopt a annealing schedule in \cite{magnet}.
That is, given the current decoding position $i \in \{ 1, ..., n \}$ and an initial CFG scale $\omega_0 \geq 1$, we gradually decrease the guidance scale by:

\begin{equation}
\omega_i = 1 + (\omega_0 - 1) \times \left(1 - \frac{i - 1}{n - 1}\right)
\label{eq:cfg_schedule}
\end{equation}

Intuitively, this means that the initial decoded tokens are more tailored to the given condition and gradually allow for greater diversity and uncertainty.
We find that this annealing schedule works better than the constant schedule and the linear schedule used in MAR~\cite{li2024autoregressive}.

\paragraph{Sampling temperature.}

Conventionally, the higher sampling temperature corresponds to more diverse samples.
To facilitate this behavior, we can multiply the noise scale $\delta$ in equation~\ref{eq:sampling} by a temperature factor $\tau$~\cite{dhariwal2021diffusion}.

\paragraph{Results.}

\begin{figure}[ht]
\vskip 0.1in
\begin{center}
\centerline{\includegraphics[width=0.6\columnwidth]{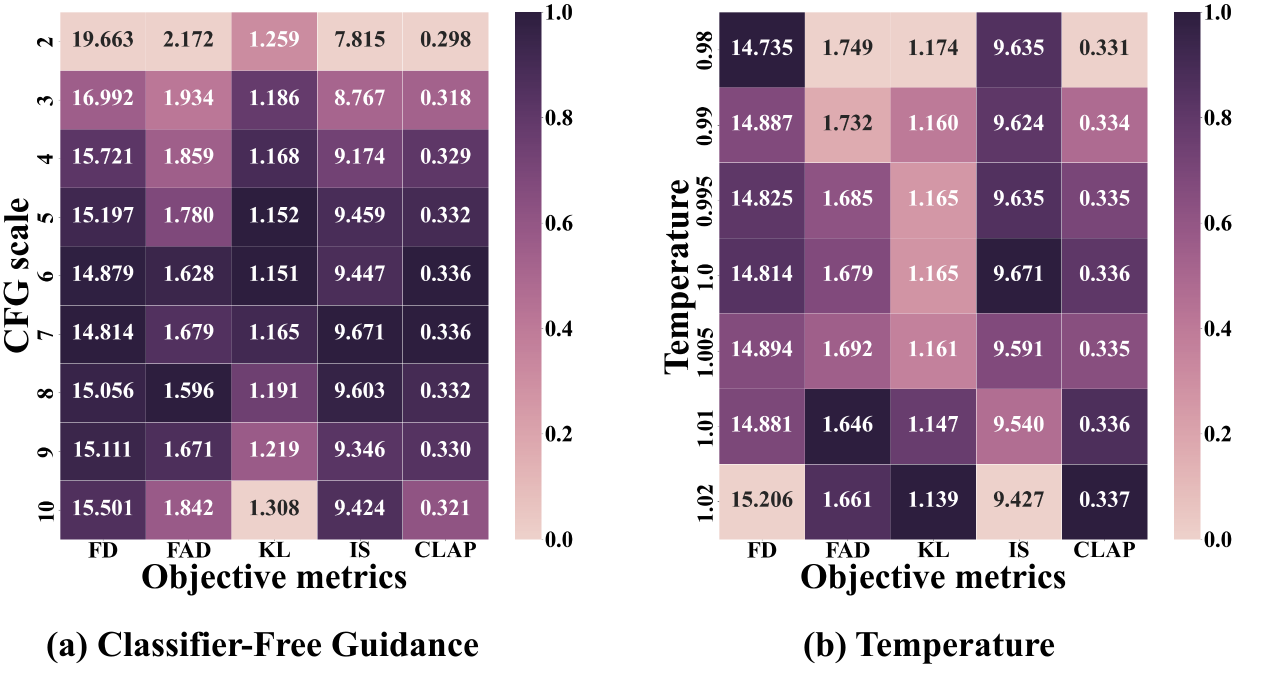}}
\caption{Ablating the (a) classifier-free guidance scale (CFG) and the (b) temperature during the MLP diffusion sampling. The darker color represents the better performance. We use the AudioMNTP Base configuration for the ablation. In (a), the temperature is fixed to 1.0; In (b) the CFG is fixed to 7.0, as suggested by (a).}
\label{image:cfg_temp}
\end{center}
\vskip -0.2in
\end{figure}

Figure~\ref{image:cfg_temp} demonstrates the effects of the CFG guidance scale $\omega_0$ and the sampling temperature $\tau$.
Figure~\ref{image:cfg_temp} (a) shows that $\omega_0$ is critical for the competitive performances, and the best performance is achieved around 7.
As a result, we set $\omega_0 = 7$ as default thorough the article.
Next, we explore different sampling temperatures with $\omega_0 = 7$ in Figure~\ref{image:cfg_temp} (b). Figure~\ref{image:cfg_temp} (b) shows that sampling temperatures have varying impacts on different metrics. As a result, given a specific metric to optimize, it is helpful to tune the temperature. However, no single $\tau$ achieves the best performance across all metrics. Therefore, we set $\tau = 1$ as the default.

\section{MNTP v.s. MLM v.s. Next-Token Prediction}
\label{appendix:mntp_vs_mlm_ntp}

We compare MNTP to the closely related MLM and next-token prediction for their similarities and differences.

\subsection{Comparing MNTP to MLM}
\label{appendix:mntp_vs_mlm}

\begin{figure}[ht]
\vskip 0.1in
\begin{center}
\centerline{\includegraphics[width=0.72\columnwidth]{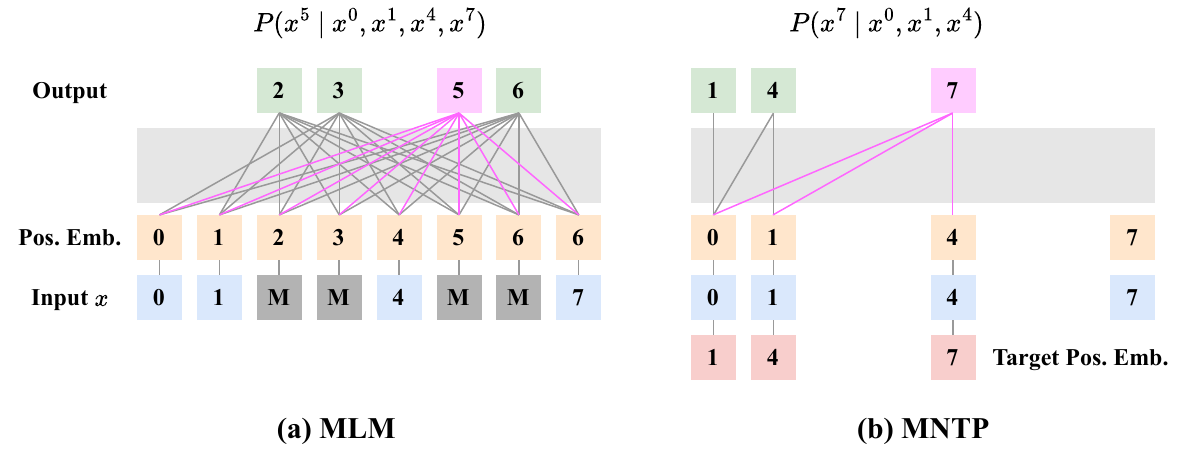}}
\caption{Comparing the differences between MLM and MNTP.}
\label{image:mlm_vs_mntp}
\end{center}
\vskip -0.1in
\end{figure}

MNTP draws inspiration from MLM to improve next-token prediction.
They share the same spirit in \textit{learning the contextualized dependencies}, where the model generates the representation from a sparse context which is predictive to the unseen data.
Figure~\ref{image:mlm_vs_mntp} illustrate the idea.
In MLM, the model predicts the unseen $x_5$ conditioning on a masked sparse context $\{ x^0, x^1, x^4, x^7 \}$.
In MNTP, the model predicts the unseen $x^7$ also conditioning on a dropped sparse context $\{ x^0, x^1, x^4 \}$.
The differences include the following:

\begin{enumerate}
    \item \textbf{Directionality}: MLM relies on a bidirectional model, while MNTP is designed for the causal model. As a result, the sparse context of MLM includes both the past $\{ x^0, x^1, x^4 \}$ and future $\{ x^7 \}$ context, while that of MNTP includes only the past context $\{ x^0, x^1, x^4 \}$.

    \item \textbf{Mask token \& Target positional embedding}: We remove the mask tokens which do not appear in the NTP decoding stage and cost extra computation, and predict the unseen tokens directly at the position of the seen tokens with the additional target positional information.

    \item \textbf{Prediction target}: We consider the right-masked span in Figure~\ref{image:mlm_vs_mntp} as an example. In MLM, all tokens in the masked span are predicted using both left and right context, e.g., \(p(x^5 \mid x^4, \ldots)\) and \(p(x^6 \mid x^7, \ldots)\). In contrast, MNTP predicts only the rightmost token using left context alone, e.g., \(p(x^7 \mid x^4, \ldots)\). To quantify task difficulty, we only list the closest observed token to the unseen token as the condition. As a result, MNTP is uni-directional and relies on more distant tokens on average compared to MLM, making it intuitively more challenging.
\end{enumerate}

\subsection{Comparing MNTP to Next-Token Prediction.}
\label{appendix:mntp_vs_ntp}

\begin{figure}[ht]
\vskip 0.1in
\begin{center}
\centerline{\includegraphics[width=0.45\columnwidth]{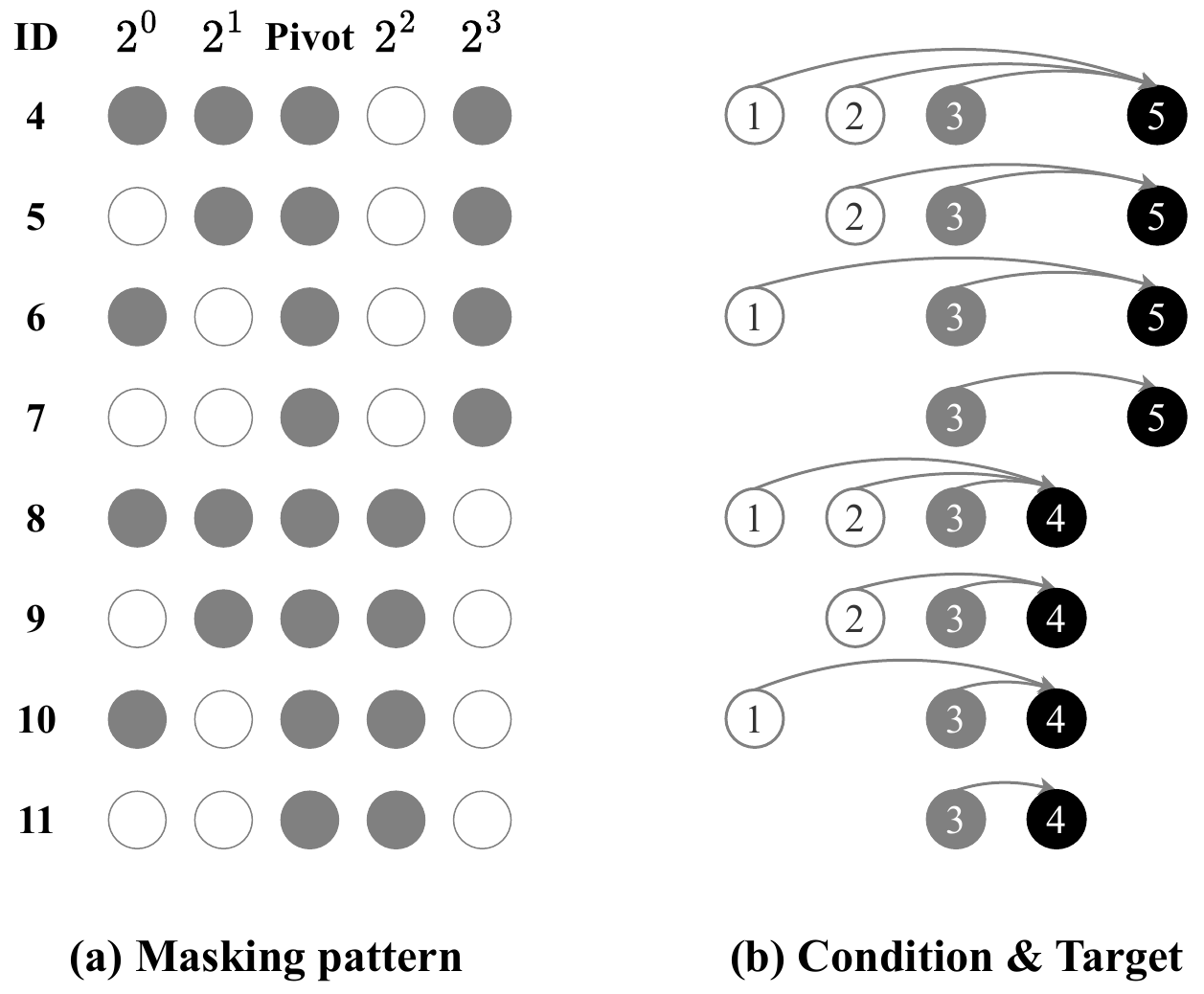}}
\caption{Visualizing the generalized causal language model. (a) shows the masking patterns. We enumerate the masking patterns with binary numbers where a masked position is denoted by 1 (white) and a unmasked position is denoted by 0 (gray). (b) shows the corresponding dropped sequence and the prediction pattern given the current gray token $x^3$. The conditions are in white and the prediction target is in black. }
\label{image:glm}
\end{center}
\vskip -0.1in
\end{figure}

We can view MNTP as a Generalized form of Causal Language Modeling (GCLM), where the model \textit{predicts any future timestamp given any subset of past information}.
We illustrate the idea with an simple example in Figure~\ref{image:glm}.
Given the length-5 sequence and the current token $x^3$, all the possible prediction patterns in GCLM are enumerated in Figure~\ref{image:glm} (b), including possible subsets of the past conditions and the possible future prediction targets.
All the corresponding masking patterns are listed in Figure~\ref{image:glm} (a), and they are all possible masking patterns in our masking schedule since we sample the masking ratio from the entire $[0, 1]$.
To be formal, we follow the notation in Section~\ref{section:AudioNTP} and Section~\ref{section:AudioMNTP}.
Given an unmasked sequence $x = \{ x^1, ..., x^n \}$, without loss of generality, we pick a current input token $x^i$ where $1 < i < n$.
We then have the past token set $P = \{x^1, ..., x^{i-1}\}$ and future token set $F = \{x^{i+1}, ..., x^{n}\}$.
For any prediction pattern in GCLM, there is a past token subset $\bar{P} \subseteq P$ and a target future token $x^f \in F$, we can derive at least one masking pattern $v = \{ v^1, ..., v^n \}$ which satisfies this prediction pattern by:

\begin{equation}
v^j = 
\begin{cases}
1 & \text{if } j \in \{i,f\} \text{ or } (j < i \text{ and } x^j \in \bar{P}), \\
0 & \text{otherwise}.
\end{cases}
\end{equation}

Note that this is not the only masking pattern satisfying this prediction pattern, our goal is to show that the prediction pattern would be sampled and learned.
This masking pattern can be sampled when setting the masking ratio $r = 1 - {\sum_{j=1}^n \frac{v^j}{n}}$.
Since our masking schedule is a distribution over the entire $[0, 1]$ (See Appendix~\ref{appendix:mask_schedule_define}), the pattern would be sampled and learned. 
As a result, our MNTP at position $i$ models $p_i(x^f~ |~ \bar{P})$ over all $x^f \in F$ and $\bar{P} \subseteq P$. 
By setting $\bar{P} = P$ and $x^f = x^{i+1}$, we restores the original next-token prediction with the full context.

\section{Ablating the components of MNTP}
\label{appendix:ablate_mntp}

\begin{figure}[ht]
\vskip 0.2in
\begin{center}
\centerline{\includegraphics[width=1\columnwidth]{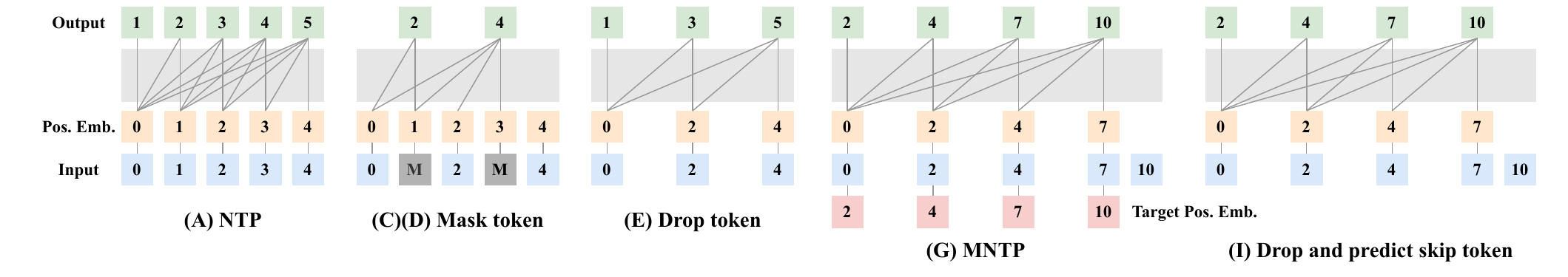}}
\caption{Visualizing the differences between the ablation variants. The subfigure IDs match the IDs in Table~\ref{table:ablate_mntp_results}.}
\label{image:mntp_ablation}
\end{center}
\vskip -0.2in
\end{figure}

We visualize the process of MNTP ablation in Figure~\ref{image:mntp_ablation}.
Refer to Section~\ref{section:mntp_ablation} for the discussion.

\section{Implementation}

\subsection{Continuous-valued audio tokenizer}
\label{appendix:audio_tokenizer}

\paragraph{Tokenize.}

We leverage the pre-processing pipeline of AudioLDM~\cite{liu2023audioldm}, including a pre-trained variational autoencoder (VAE)~\cite{Kingma2014} $V = \{ V_E, V_D \}$ and a pre-trained Hifi-GAN~\cite{kong2020hifi} vocoder $G$.
$V_E$ and $V_D$ are the encoder and the decoder, respectively.
We first extract the 64-band Mel-spectrogram with the hop size of 10 ms from audio: $m \in R^{T \times F}$, where $T = s \cdot 1000 / 10$ and $F = 64$.
Given that all our training samples are 16 kHz 10-second audios, $s = 160000$.
The VAE encoder $V_E$ then encodes $m$ into a 2-D map of continuous latents $v \in R^{\frac{T}{r} \times \frac{F}{r} \times c}$ where $r=4$ is the compression level and $c=8$ is the latent dimension.
To reduce the sequence length and the training cost, we further patchify $v$ into $\bar{x} \in R^{\frac{T}{r \cdot p} \times \frac{F}{r \cdot p} \times (c \cdot p^2)}$ by stacking the latents in a $p \times p$ neighborhood.
We set $p=4$. Finally, we flatten $\bar{x}$ into a 1-D sequence $x = \{ x^1, ..., x^n \}$ in a row-major order, where $n = \frac{T}{r \cdot p} \times \frac{F}{r \cdot p}$ and $x^i \in R^{c \cdot p^2}$.
That is, different frequency components are placed together in the flatten sequence.
With the tokenization, we represent a 10-second clip by 256 continuous-valued tokens with the latent size 128.

\paragraph{De-tokenize.} Given the sampled, flattened tokens $\tilde{x} = \{ \tilde{x}^1, ..., \tilde{x}^n \}$, we undergo the reverse process of flatten and patchify to get the 2-D latent map $\tilde{v} \in R^{{\frac{T}{r} \times \frac{F}{r} \times c}}$.
The VAE decoder $V_D$ then decodes $\tilde{v}$ into the sampled Mel-spectrogram $\tilde{m}$.
Finally, the vocoder $G$ synthesizes waveforms $\tilde{a}$ from $\tilde{m}$.

\subsection{Conditional audio language modeling}
\label{appendix:continuous_language_modeling}

\paragraph{Diffusion head.}

We follow \cite{li2024autoregressive} for its diffusion process, including the cosine noise schedule, training/sampling (1000/100) diffusion steps, and the prediction target (noise).
We reuse the MLP architecture introduced in \cite{li2024autoregressive}.
The MLP of the Base model has 3 layer, and that of the Large model has 6 layer.
Both versions use 1024 dimension for all the MLP layers.

\paragraph{Transformer.}

We use the Transformer~\cite{vaswani2017attention} implementation in ViT~\cite{wang2021not}, same as MAR~\cite{li2024autoregressive}.
The Base model adopts 24 layers with the hidden dimension of 768; the Large model adopts 32 layers with the hidden dimension of 1024.
The model sizes are about 170M and 420M, respectively.
Since we use the same architecture, we can leverage the image pre-trained weights in MAR.
The image pre-trained weights help the spectrogram-based audio models, as suggested by AST~\cite{gong21b_interspeech}.
We reuse only the weights of the backbone Transformer, excluding those of the diffusion head, since the latter is entangled to the modality-specific token dimension.

\paragraph{Text prompt.}

We use CLAP~\cite{wu2023large} and FLAN-T5~\cite{chung2024scaling} to extract the text embeddings and concatenate them as the input prompt.
Our ablation study shows that using both is slightly better than using either individually.
In practice, FLAN-T5 produces variable-length embeddings, which we truncate to the first 77 embeddings. Combined with a single CLAP embedding and the linear projections, our text conditioning is equal to 78 tokens, which we use as the initial input prompt in the causal LM for audio generation. The positions of the text prompt do not contribute to the loss.

\section{Evaluation}

\subsection{Objective evaluation}
\label{appendix:objective_eval}

We evaluate our model on the AC evaluation set.
Each audio is labeled by 5 captions, and we follow
AudioLDM to randomly select one text description as the input prompt.
We leverage the AudioLDM evaluation toolkit\footnote{
\href{https://github.com/haoheliu/audioldm_eval}{https://github.com/haoheliu/audioldm\_eval}
}.
Given the generated audio and the ground-truth audio in AC, we calculate several metrics: Fréchet Audio Distance (FAD)~\cite{kilgour19_interspeech}, Fréchet Distance (FD), Kullback--Leibler divergence (KL), Inception Score (IS)~\cite{liu2023audioldm}, and Contrastive Language-Audio Pretraining (CLAP) score~\cite{huang2023make}.
Intuitively, lower FD and FAD indicate higher similarity to the paired ground-truth. The lower KL indicates that the set of the generated audios share the similar distribution as the ground-truth audios. The higher IS indicates that given an audio classifier, each audio is well-recognized by the classifier\footnote{Indicated by the low entropy of the output distribution.}, and the classifier classifies all the generated audios into diverse classes\footnote{Indicated by the high entropy of the average output distribution over the generated audios.}.
Intuitively, these two conditions jointly imply the generated audios are diverse and authentic.
Finally, the higher CLAP score indicates the better alignment between the generated audio and the input text prompt.

\subsection{Subjective evaluation}
\label{appendix:subjective_eval}

We follow a similar approach to Tango~\cite{tango}, we assessed 20 generated audio samples based on text relevance (REL) and overall quality (OVL) but used a 1-to-5 rating scale instead of a 1-to-100 scale.
Each sample was rated by at least 10 participants.
We separate the evaluation into speech and non-speech categories to understand the differences in model behaviors.
The speech prompts are sampled by selecting the text prompts containing man, woman, and person and people.
The non-speech prompts are sampled from the remaining.
Note that in the AudioCaps evaluation set, most prompts consist of both speech and non-speech descriptions.
We compare our method to the existing decoder-only solution AudioGen and two SOTA diffusion models AudioLDM 2 and Tango 2.

%% file: icml2024/tables/datasets.tex
\begin{table*}[h]
\caption{\textbf{The datasets used by the TTA systems.} The AS, AC, WC stand for AudioSet~\cite{gemmeke2017audio}, AudioCaps~\cite{kim2019audiocaps}, WavCaps~\cite{mei2024wavcaps}, respectively. WavCaps is composed of FreeSound~\cite{fonseca2017freesound}, BBC Sound Effects\textsuperscript{\ref{bbc}}, SoundBible\textsuperscript{\ref{soundbible}}, and the AudioSet Strongly-Labeled Subset~\cite{hershey2021benefit}, augmented with the ChatGPT-generated text prompt. Some systems use the individual datasets in WavCaps without the natural language prompt, i.e. AudioGen.
}
\label{bg_internal_main}
\vskip 0.05in
\centering
\begin{sc}
\resizebox{\linewidth}{!}{
    \begin{tabular}{lc}
    \toprule

    Method & Datasets \\

    \midrule
    \midrule

    \makecell[l]{AudioGen Base~\cite{AudioGen} \\ AudioGen Large~\cite{AudioGen} \\ MAGNET-Small (Audio\textsuperscript{\ref{magnet}})~\cite{magnet} \\ MAGNET-Large (Audio)~\cite{magnet}} & 
    \makecell{\textbf{AS} + \textbf{AC} + BBC sound effects + Clotho v2~\cite{drossos2020clotho} \\ + VGG-Sound~\cite{chen2020vggsound} + FSD50K~\cite{fonseca2021fsd50k} \\ + Free To Use Sounds\textsuperscript{\ref{freetousesounds}} + Sonniss Game Effects\textsuperscript{\ref{gameaudio}} + WeSoundEffects\textsuperscript{\ref{wesoundeffects}} \\ + Odeon Sound Effects\textsuperscript{\ref{paramountmotion}}} \\

    \midrule

    Tango~\cite{tango} & {\textbf{AC}} \\

    \midrule

    Tango-Full-FT~\cite{tango} & \makecell{
    \textbf{AS} + \textbf{AC} + \textbf{WC} + MusicCaps~\cite{agostinelli2023musiclm} \\ + ESC~\cite{piczak2015esc} + UrbanSound~\cite{salamon2014dataset} \\ + GTZAN~\cite{tzanetakis2002musical} + Musical Instruments\textsuperscript{\ref{music_instrument}}
    } \\

    \midrule

    Tango-AF\&AC-FT~\cite{kong2024improving} & \textbf{AC} + AF-AudioSet\textsuperscript{\ref{af-audioset}} \\

    \midrule

    Tango 2~\cite{tango2} & Tango-Full-FT data + Audio-Alpaca (AA)\textsuperscript{\ref{audio-alpaca}} \\

    \midrule

    Make-an-Audio 2~\cite{make-an-audio2} & \makecell{\textbf{AS} + \textbf{AC} + \textbf{WC} + WavText5K~\cite{deshmukh23_interspeech} + Adobe Audition Sound Effects\textsuperscript{\ref{adobeauditiondlcsfx}} \\ 
    + MACS~\cite{martin2021ground} + Clothv2~\cite{drossos2020clotho} \\ 
    + Audiostock\textsuperscript{\ref{audiostock}} + Epidemic Sound\textsuperscript{\ref{epidemicsound}} + FSD50K~\cite{fonseca2021fsd50k} + Odeon Sound Effects\textsuperscript{\ref{paramountmotion}} \\
    UrbanSound~\cite{salamon2014dataset} + ESC~\cite{piczak2015esc} + TUT~\cite{mesaros2016tut}
    } \\

    \midrule

    \makecell[l]{AudioLDM2-AC~\cite{AudioLDM2} \\ AudioLDM2-AC-Large~\cite{AudioLDM2}} & {\textbf{AC}} \\

    \midrule

    \makecell[l]{AudioLDM2-Full~\cite{AudioLDM2} \\ AudioLDM2-Full-Large~\cite{AudioLDM2}} & {\textbf{AS} + \textbf{AC} + \textbf{WC} + VGG-Sound~\cite{chen2020vggsound} + MSD~\cite{Bertin-Mahieux2011}} \\

    \midrule

    UniAudio~\cite{uniaudio} & \makecell{\textbf{AS} + \textbf{AC} + \textbf{WC} + LibriLight~\cite{kahn2020libri} + LibriTTS~\cite{zen19_interspeech} \\ + MLS~\cite{pratap20_interspeech} + AISHELL3~\cite{shi21c_interspeech} + OpenCPOP~\cite{wang22b_interspeech} \\ + OpenSinger~\cite{huang2021multi} + MSD~\cite{Bertin-Mahieux2011} \\ + PromptSpeech~\cite{guo2023prompttts} + openSLR26,openSLR28F~\cite{ko2017study}} \\

    \midrule

    \makecell[l]{\textbf{Audio-NTP Base} \\ \textbf{Audio-MNTP Base} \\ \textbf{Audio-MNTP Large}} & {\textbf{AC} + \textbf{WC}} \\

    \bottomrule
    \end{tabular}
}
\end{sc}
% \end{small}
\vskip -0.1in
\end{table*}

\setcounter{footnote}{18}
\footnotetext{\label{bbc}https://sound-effects.bbcrewind.co.uk}

\addtocounter{footnote}{1}
\footnotetext{\label{soundbible}https://soundbible.com/}

\addtocounter{footnote}{1}
\footnotetext{\label{magnet}There are music and audio versions of Magnet. The music version follows the datasets used in \cite{copet2024simple} and the audio version follows those in AudioGen.}

\addtocounter{footnote}{1}
\footnotetext{\label{freetousesounds}https://www.freetousesounds.com/all-in-one-bundle/}

\addtocounter{footnote}{1}
\footnotetext{\label{gameaudio}https://sonniss.com/gameaudiogdc}

\addtocounter{footnote}{1}
\footnotetext{\label{wesoundeffects}https://wesoundeffects.com/we-sound-effects-bundle-2020/}

\addtocounter{footnote}{1}
\footnotetext{\label{paramountmotion}https://www.paramountmotion.com/odeon-sound-effects}

\addtocounter{footnote}{1}
\footnotetext{\label{music_instrument}https://www.kaggle.com/datasets/soumendraprasad/musical-instruments-sound-dataset}

\addtocounter{footnote}{1}
\footnotetext{\label{af-audioset}AF-AudioSet is a synthetic dataset released by \cite{kong2024improving}, where Audio Flamingo~\cite{kong2024audio} provides the caption for the AudioSet audio.}

\addtocounter{footnote}{1}
\footnotetext{\label{audio-alpaca}The Audio-Alpaca (AA) preference dataset is released by Tango 2, where a text prompt corresponds to a winner audio and a loser audio. It is used after the standard TTA supervised training for the preference alignment.}

\addtocounter{footnote}{1}
\footnotetext{\label{adobeauditiondlcsfx}https://www.adobe.com/products/audition/offers/adobeauditiondlcsfx.html}

\addtocounter{footnote}{1}
\footnotetext{\label{audiostock}https://audiostock.net/}

\addtocounter{footnote}{1}
\footnotetext{\label{epidemicsound}https://www.epidemicsound.com/}

%% file: icml2024/tables/mask_schedule.tex
\begin{table}[h]
\caption{Ablating the masking schedule of MNTP. The schedule IDs match the IDs in Figure~\ref{image:masking_schedule}.}
\label{table:mntp_schedule}
\vskip 0.05in
\centering
\begin{sc}
\resizebox{0.45\linewidth}{!}{
    \begin{tabular}{l|ccccc}
    \toprule
    Schedule ID & FD $\downarrow$ & FAD $\downarrow$ & KL $\downarrow$ & IS $\uparrow$ & CLAP $\uparrow$ \\

    \midrule

    (A) & 15.22 & 1.70 & 1.17 & 8.83 & 0.326 \\
    (B) & 15.96 & 2.17 & 1.25 & 8.18 & 0.293 \\
    (C) & \textbf{14.81} & \textbf{1.68} & \textbf{1.16} & \textbf{9.67} & \textbf{0.336} \\
    (D) & 15.55 & 1.89 & 1.16 & 9.56 & 0.327 \\

    \bottomrule
    \end{tabular}
}
\end{sc}
% \end{small}
\vskip -0.1in
\end{table}